\documentclass[12pt]{iopart}
\usepackage{iopams,xcolor}  
\usepackage[percent]{overpic}

\expandafter\let\csname equation*\endcsname\relax
\expandafter\let\csname endequation*\endcsname\relax
\usepackage{amsmath}

\begin{document}

\newcommand{\bra}[1]    {\langle #1|}
\newcommand{\ket}[1]    {|#1 \rangle}
\newcommand{\ketbra}[2]{|#1\rangle\!\langle#2|}
\newcommand{\braket}[2]{\langle #1 | #2\rangle}
\renewcommand{\tr}[1]    {{\rm Tr}\left[ #1 \right]}
\newcommand{\trs}[1]    {{\rm Tr}\big[ #1 \big]}
\newcommand{\av}[1]    {\langle #1 \rangle}
\newcommand{\avs}[1]    {\langle #1 \rangle}
\renewcommand{\mod}[1]    {\left| #1 \right|}
\newcommand{\modsq}[1]    {\left| #1 \right|^2}
\newcommand{\modsqs}[1]    {\big| #1 \big|^2}
\newcommand{\rk}{\text{rk}}

\newcommand{\En}{\mathcal E_L}
\newcommand{\Q}{\mathcal Q}
\newcommand{\Qe}{\mathcal Q_{\rm ent}}
\newcommand{\Qb}{\mathcal Q_{\rm Bell}}
\newcommand{\Qep}[1]{\mathcal Q_{\rm ent}^{(p=#1)}}
\newcommand{\Qet}{\tilde{\mathcal Q}_{\rm ent}}
\newcommand{\Qbt}{\tilde{\mathcal Q}_{\rm Bell}}

\newcommand{\cor}[1]{\log_4\left(4^N\modsqs{\avs{#1}}\right)}
\newcommand{\cz}{\hat C_z}

\newcommand{\Enq}{\mathcal E_{L}^{(q)}}

\newcommand{\fs}{\frac1{\sqrt2}}

\newcommand{\cohb}[1]{\mathcal C_{ghz}^{(#1)}}
\newcommand{\coh}{\mathcal C_{\textrm{GHZ}} }

\newcommand{\nd}{\textsuperscript{nd} }
\newcommand{\rd}{\textsuperscript{rd} }
\renewcommand{\th}{\textsuperscript{th} }

\newcommand{\red}[1]{\textcolor{red}{#1}}

\title{Many-body quantum resources of  graph states}

\author{Marcin Płodzień$^{1}$, Maciej Lewenstein$^{1,2}$ and Jan Chwedeńczuk$^{3}$}

\address{  $^{1}$ICFO-Institut de Ciencies Fotoniques, The Barcelona Institute of Science and Technology, 08860 Castelldefels (Barcelona), Spain\\
  $^{2}$ICREA, Passeig Lluis Companys 23, 08010 Barcelona, Spain\\
  $^{3}$ Faculty of Physics, University of Warsaw, ul. Pasteura 5, 02-093 Warszawa, Poland}
\vspace{10pt}

\begin{abstract}
  Characterizing the non-classical correlations of a complex many-body system is an important part of quantum technologies. An ideal tool for this task would scale well with the size of the 
  system, be easily computable and be easily measurable.
  In this work, we focus on graph states, which are promising platforms for quantum computation, simulation, and metrology.
  We consider four topologies: star graph states with edges, Tur\'an graphs, $r$-ary tree graphs, and square grid cluster states.
  We provide a method to characterize their quantum content:  
  many-body Bell correlations, non-separability and entanglement strength for an arbitrary number of qubits. 
  We also relate the strength of these correlations to the usefulness of graph states for quantum sensing. Finally, we characterize many-body entanglement in graph states with up 
  to eight qubits in $146$ classes that are not equivalent under local transformations or graph isomorphisms. 
  This technique is straightforward and does not require any assumptions about the multi-qubit state; therefore it could be applied wherever precise knowledge of many-body quantum correlations is necessary.
\end{abstract}
 
\section{Introduction} 
Quantum coherence and many-body quantum correlations---such as entanglement and Bell correlations--- 
are key features of many-body quantum systems~\cite{RevModPhys.89.041003,horodecki2009entanglement,brunner2014bell,Srivastava2024,horodecki2024multipartiteentanglement}. 
In addition to being fundamental aspects, they are also considered vital resources
for future quantum-based technologies~\cite{RevModPhys.91.025001,https://doi.org/10.48550/arxiv.2405.05785}. 
In order to trigger the ``second quantum revolution'', we must learn to prepare and store the many-body quantum states in a controlled way, and subsequently characterize and certify their quantum content.

\begin{figure}[t!]
\centering
  \begin{overpic}[width=0.8\linewidth]{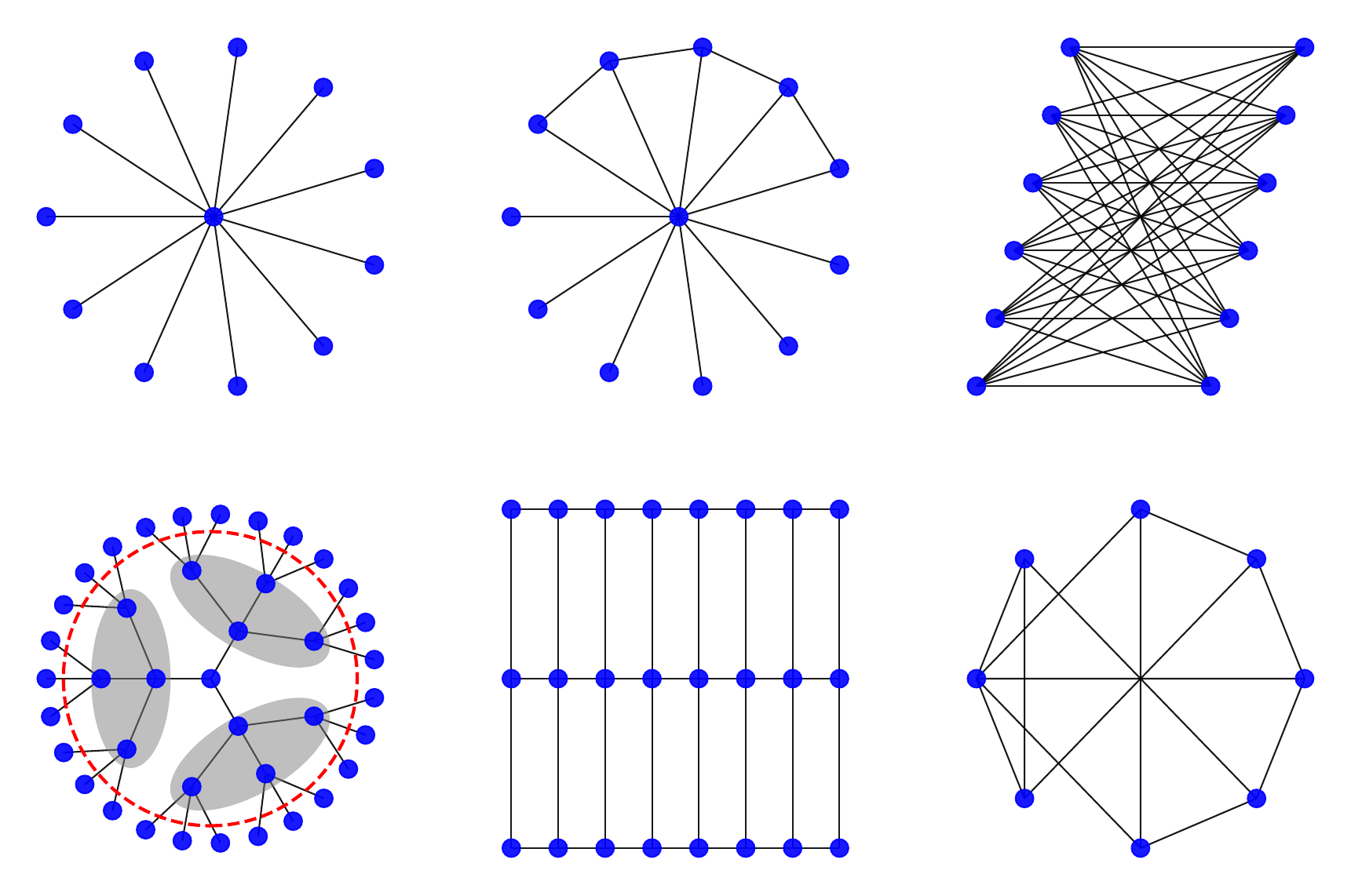}
    \put(5,65){\makebox(0,0)[c]{(a)}}
    \put(40,65){\makebox(0,0)[c]{(b)}}
    \put(70,65){\makebox(0,0)[c]{(c)}}
    \put(5,32){\makebox(0,0)[c]{(d)}}
    \put(40,32){\makebox(0,0)[c]{(e)}}
    \put(70,32){\makebox(0,0)[c]{(f)}}
  \end{overpic}
  \caption{Examples of distinct graph state topologies. (a) star graph with $L=12$ nodes, (b) star graph with $L=12$ nodes and $p=4$ edges, (c) Tur\'an graph with $L = 12$ nodes partitioned into equal 
    $K = 2$ groups, (d) $r$-ary tree with $r=3$ roots and $h=2$ depth.The shaded areas 
      show separate star graphs forming GHZ states, which is a useful observation for computation
    of the Bell correlator. The dashed red circle shows that the tree graph can be cut into closed layers formed by a fractal-like hierarchy of star graphs (see Section~\ref{sec.trees}).} 
    Panel (e) shows a square grid graph and panel (f) shows an eight-qubit graph state representative of one of $146$ non-equivalent classes under local unitary transformations and graph isomorphisms.
  \label{fig:fig1}
\end{figure}

In this context, an important family of many-body quantum systems is that of graph states, in which qubits are placed at nodes and connected by edges that represent two-body interactions. 
Graphs can have different topologies that resemble stars [Fig.~\ref{fig:fig1}(a)], stars with edges [Fig.~\ref{fig:fig1}(b)], Tur\'an graphs where the 
nodes are partitioned into groups that are not connected with each other [Fig.~\ref{fig:fig1}(c)], or tree-like fractals [Fig.~\ref{fig:fig1}(d)].
Some topologies, such as cluster graphs, see Fig.~\ref{fig:fig1}(e), enable universal quantum computation through a series of local measurements and classical feed-forward operations, 
known as measurement-based quantum computing~\cite{jozsa2006introduction,PhysRevLett.86.5188,PhysRevA.68.022312, doi:10.1142/S0219749904000055,PhysRevA.68.022312,
  hein2006entanglement,PhysRevLett.98.220503,briegel2009measurement,PhysRevLett.98.220503,Raza2023}. 
Graph states have been  studied as a resource for quantum communication~\cite{PhysRevA.86.042304,Bell2014,PhysRevA.100.052333,Javelle2013,Hahn2019,Hilaire2021,Ding2023,Fischer2021, Unnikrishnan2022,PhysRevA.108.012402,https://doi.org/10.48550/arxiv.2407.01429,Makuta2023,li2024}, 
and more recently for quantum metrology~\cite{PhysRevLett.124.110502}. Each graph state belongs to an equivalence class under local Clifford operations \cite{BOUCHET199375,PhysRevA.69.022316,Adcock2020}, 
so preparing graph states optimally relates to finding the state orbit with the fewest qubit connections~\cite{PhysRevA.83.042314}.
They have been prepared experimentally in various physical platforms, including photonic systems~\cite{Russo2019,Thomas2024,PRXQuantum.5.020346}, spin-1/2 systems~\cite{PhysRevA.109.042604}, 
atomic ensembles~\cite{Clark2005,PhysRevA.83.010302,Cooper2024}, and solid state systems, such as quantum dots~\cite{Li2020} and superconducting qubtis~\cite{Gnatenko2021}. 

Future applications require that we characterize the ``quantum content'' of graph states: their strength and how they scale with system size.
Once the quantum content  of a graph state is known, we should be able to determine how correlations change when edges are added or removed.
The entanglement of graph states is typically characterized by the Schmidt measure, geometric measures ~\cite{Markham2007,PhysRevA.64.022306, Guo_2014,10.1063/1.4903126}, 
or the projector-based entanglement witness~\cite{PhysRevLett.94.060501,zander2024}. The former defines the degree of entanglement as the supremum of 
the fidelity between the given state and all product states. The latter refers to the overlap between the given state and the Greenberger-Horne-Zeilinger (GHZ) state. 
These entanglement measures have greatly expanded our understanding of graph states. However, they are not always easy to use. They can be difficult 
to calculate numerically, and they are often applicable only to a certain set of states. 

In this work, we propose a new method for characterizing the quantum content of graph states that satisfies the following criteria: 
(i) it does not rely on distance in the Hilbert space so it is easily computable;
(ii) it directly shows how quantum correlations change when graph connectivity changes;
(iii) it is experimentally accessible; 
(iv) it can be applied to both pure and mixed states, which is important when coupling gates are imperfect.

This method only requires knowledge of one element of the  graph state's density matrix. From this element, one can infer the depth of entanglement and  many-body Bell correlations
~\cite{zukowski2002bell,mermin1990extreme,10.1119/1.12594, cavalcanti2007bell, Reid2012, cavalcanti2011unified,Ardehali_1992}.
The method is scalable in two ways. First, it applies to graph states of arbitrary size. Second, it enables a systematic study of how correlation strength
changes when interactions between qubits (i.e., edges) are modified. This provides control over the quantum resources needed for a given task. 
Additionally, the aforementioned matrix element used by this method can be accessed experimentally by measuring
of the multiple quantum coherences~\cite{garttner2017measuring,PhysRevLett.120.040402}.

We discuss the application of the proposed method to four graph topologies: star graphs, Tur\'an graphs, $r$-ary tree graphs, and cluster states~\cite{book_graphs}.
We also discuss the application to $146$ non-equivalent classes of graph states with up to eight qubits with respect to local unitary (LU) transformations and graph isomorphisms.
We provide analytical results for the amount of quantum correlation for any number of qubits. Finally, we demonstrate the quantum-enhanced metrological utility of these graphs.

\section{Preliminaries}
This opening section presents some of the basic tools and concepts that will be used throughout the text. 

\subsection{Graph states}
A graph $G$   is a pair $G = (V,E)$, consisting of a set  of $|V|$ nodes and a set of $|E|$ edges connecting any two nodes where  $(k, l) \in E$, and $k,l\in V$. 
A graph state denoted by $\ket{\psi}$ is defined on a graph $G$ as follows: each node $v_k\in V$ corresponds to a qubit in an eigenstate of $\hat{\sigma}_x$ 
operator with $+1$ eigenvalue, namely
\begin{align}
  \hat\sigma_x^{(k)}\ket{0^{(k)}_x}=+1\ket{0^{(k)}_x},
\end{align}
giving the initial state in the form
\begin{align}
  \ket{\psi_{\rm ini}} =  \bigotimes_{k=1}^L \ket{0^{(k)}_x},
\end{align}
where $L=|V|$ is the number of nodes. Next, edge $(k,l)$ corresponds to the action of a Control-Z gate $\hat{C}_z^{(kl)}$ between nodes $k$ and $l$, defined as
\begin{align}
  \hat C^{(kl)}_z=\ketbra{00}{00}_z+\ketbra{01}{01}_z+\ketbra{10}{10}_z-\ketbra{11}{11}_z,
\end{align}
with $\ket{00}_z=\ket{0_z^{(k)}}\otimes\ket{0_z^{(l)}}$ {\it etc}. Therefore the graph state is
\begin{align}
  \ket\psi = \prod_{(k,l)\in E(G)}\hat{C}_{z}^{(kl)}\ket{\psi_{\rm ini}}.
\end{align}
The quantum correlations that we investigate in this work stem from interactions that are represented by a pattern of edges.

\subsection{Bell correlator}

We use the following many-body correlator to certify the quantum content of a given graph state~\cite{spiny.milosz,PhysRevA.102.013328,PhysRevLett.126.210506,PhysRevLett.129.250402,PhysRevA.107.013311,PhysRevB.108.104301,PhysRevResearch.6.023050,PhysRevA.110.012210,grze_mar,plodzien2024inherent}. 
Consider a system of $L$ parties each of which provides two binary results
\begin{align}\label{eq.binary}
  \sigma^{(k)}_x=\pm1,\ \ \ \sigma^{(k)}_y=\pm1,\ \ \ \ \ \ (k\in\{1\ldots L\}).
\end{align}
These outcomes are multiplied to construct the  correlation function
\begin{align}\label{eq.corr.gen}
  \En=\modsqs{\av{\prod_{k=1}^L\sigma_+^{(k)}}}.
\end{align}
The average is taken over the ensemble of experimental realizations and
\begin{align}\label{eq.rising}
  \sigma_{+}^{(k)}=\frac12(\sigma_x^{(k)}+i\sigma_y^{(k)}). 
\end{align}
If the mean in Eq.~\eqref{eq.corr.gen} is consistent with the postulates of local realism, then it can be expressed as follows:
\begin{align}\label{eq.lhv.2}
  \av{\prod_{k=1}^L\sigma_+^{(k)}}=\int\!\!d\lambda\,p(\lambda)\prod_{k=1}^L\sigma_+^{(k)}(\lambda),
\end{align}
where $p(\lambda)$ is the probability distribution of a random variable.
Since $\modsqs{\sigma_+^{(k)}(\lambda)}=1/2$ [see Eqs~\eqref{eq.binary} and~\eqref{eq.rising}], the Cauchy-Schwarz inequality (CSI) for complex integrals yields
\begin{align}\label{eq.lhv}
  \En\leqslant\int\!\!d\lambda\,p(\lambda)\prod_{k=1}^L\modsqs{\sigma_+^{(k)}(\lambda)}=2^{-L}.
\end{align}
For this reason, we conclude that
\begin{align}\label{eq.bell.l}
  \En\leqslant2^{-L}
\end{align}
is the $L$-body Bell inequality, because its violation defies the postulates of local realism expressed by Eq.~\eqref{eq.lhv.2}.
When the parties are 
quantum two-level objects, $\sigma_+^{(k)}$'s are replaced with the rising Pauli operators and the average is calculated using the density matrix $\hat\varrho$, giving the $L$-qubit Bell inequality
\begin{align} 
  \Enq[\hat\varrho]=\modsqs{{\rm Tr}\bigg[{\hat{\varrho}\bigotimes_{k=1}^L\hat\sigma_+^{(k)}}\bigg]}\leqslant2^{-L}.
\end{align}
The superscript $(q)$ is added to distinguish the general expression for the correlator, see Eq.~\eqref{eq.corr.gen}, from its ($q$)uantum equivalent.
Furthermore, if the density matrix, denoted by $\hat\varrho$ is fully separable,
\begin{align}\label{eq.sep}
  \hat\varrho = \int\! d\lambda\, p(\lambda)\bigotimes_{k=1}^{L}\hat\varrho^{(k)}(\lambda),
\end{align}
then using the upper bound for a single qubit
\begin{align}\label{eq.bloch}
  \modsqs{\tr{\hat\varrho^{(k)}(\lambda)\hat\sigma_+^{(k)}}}\leqslant4^{-1},
\end{align}
which stems from the fact that it is constrained to the Bloch sphere. By applying the CSI again, we obtain
\begin{align}\label{eq.ent.l}
  \Enq[\hat\varrho]\leqslant 4^{-L}.
\end{align}
A violation of this inequality indicates the presence of entanglement in the $L$-qubit system, because it contradicts the assumption in Eq.~\eqref{eq.sep}. 
Note that if Eq.~\eqref{eq.bell.l} is violated, then so is the inequality in Eq.~\eqref{eq.ent.l}. This is
  a natural consequence of the fact that for a quantum system to be Bell-correlated it must be entangled~\cite{horodecki2009entanglement,brunner2014bell}. However, entanglement is not sufficient 
  to break the bound~\eqref{eq.bell.l}, because the derivation of Eq.~\eqref{eq.ent.l} assumes that each party is constrained to the Bloch sphere [see Eq.~\eqref{eq.bloch}]. This restriction  is absent
  in~\eqref{eq.binary}.

Note also that $\Enq$ contains the $L$-fold action of raising operators. Therefore the  correlator depends only on the modulus squared of a single element of the density 
matrix $\hat\varrho$---the GHZ coherence $\varrho_{0^{\otimes L},1^{\otimes L}}$ between vectors $\ket{0}^{\otimes L}$ and $\ket{1}^{\otimes L}$, i.e., 
\begin{align}
  \Enq=\modsq{\varrho_{0^{\otimes L},1^{\otimes L}}}.
\end{align}
Therefore,  the GHZ state
\begin{align}\label{eq.ghz}
  \ket{{\rm GHZ}}=\frac1{\sqrt2}(\ket{1}^{\otimes L}+\ket{0}^{\otimes L}),
\end{align}
gives the maximum possible value of the Bell correlator, namely
\begin{align}\label{eq.ghz.corr}
  \Enq[{\rm GHZ}]=4^{-1}.
\end{align}
Given a quantum state $\hat\varrho$, the orientation of the raising operators can be chosen independently for each spin. This allows us to adapt the correlator to the
geometry of the state $\hat\varrho$ and maximize its value.
Therefore, the general expression for $\Enq$ that we use in this work is
\begin{align}\label{eq.elq}
  \Enq[\hat\varrho]=\max_{\vec{\kappa}} \modsqs{{\rm Tr}\bigg[{\hat{\varrho}\bigotimes_{k=1}^L\hat\sigma_{+,\kappa_k}^{(k)}}\bigg]},
\end{align}
where $\vec{\kappa} = \{\kappa_1,\dots,\kappa_L\}$ is a vector of orientations of the raising operators  $\hat\sigma^{(k)}_{+,\kappa_k}$, $\kappa_k=x,y,z$, namely
\begin{subequations}
  \begin{align}
    &\hat\sigma_{+,x}^{(k)} = \frac12(\hat\sigma_{y}^{(k)}+i\hat\sigma_{z}^{(k)}),\\
    &\hat\sigma_{+,y}^{(k)} = \frac12(\hat\sigma_{z}^{(k)}+i\hat\sigma_{x}^{(k)}),\\
    &\hat\sigma_{+,z}^{(k)} = \frac12(\hat\sigma_{x}^{(k)}+i\hat\sigma_{y}^{(k)}).
  \end{align}
\end{subequations}
The optimal vector $\vec\kappa$ that maximizes $\Enq$ is denoted by $\vec{\kappa}^*$. For illustration, consider a following GHZ four-qubit state
\begin{align}
  \ket{\mathrm{GHZ}_4}=\frac1{\sqrt2}\left(\ket{0_x,0_y,0_z,0_y}+\ket{1_x,1_y,1_z,1_y}\right).
\end{align}
The optimal orientation of the four rising operators is given by the vector $\vec\kappa^*=(x,y,z,y)$ meaning that the Bell correlator is
\begin{align}
  \Enq[{\rm GHZ}]=\modsq{\bra{\mathrm{GHZ}_4}\hat\sigma_{+,x}^{(1)}\hat\sigma_{+,y}^{(2)}\hat\sigma_{+,z}^{(3)}\hat\sigma_{+,y}^{(4)}\ket{\mathrm{GHZ}_4}}=4^{-1}.
\end{align}
Other alignments would produce a smaller, and therefore non-optimal, value of $\Enq$.

\subsection{Parametrization of the correlator}

Below we use a convenient parametrization 
\begin{align}\label{eq.gamma}
  \Enq[\hat\varrho]=4^{-(1+\gamma)},
\end{align}
where the exponent $\gamma\geqslant0$ depends on the state $\hat\varrho$, and $\gamma=0$ for the GHZ state, see Eq.~\eqref{eq.ghz.corr}. 
The larger the value of $\gamma$, the weaker the many-body correlations. Hence $\gamma$ can be interpreted as the ``distance'' of $\hat\varrho$ from the GHZ state~\eqref{eq.ghz}.

The entanglement criterion and the Bell inequality are given by Eqs~\eqref{eq.bell.l} and~\eqref{eq.ent.l}, and become respectively
\begin{align}
  \gamma<\frac{L}{2}-1\ \ \ \mathrm{and}\ \ \ \gamma<L-1.
\end{align}
In addition to certifying the entanglement and Bell correlations, $\Enq$, and thus $\gamma$, carries information about their strength~\cite{spiny.milosz}. 
If  
\begin{align}\label{eq.depth.b}
  \frac{l+1}2>\gamma\geqslant\frac l2,
\end{align}
with $l\in\mathbb N$, then the system is maximally $(l+1)$-local. For example, consider $\gamma=1/2$, so that $l=1$. According to Eq.~\eqref{eq.gamma} this gives $\Enq=\frac18$. 
This can be reproduced when the entire system consists of maximally two subsystems: one with  $L-1$ particles forming a GHZ state and one with a single uncorrelated particle yielding binary outcomes, 
see Eq.~\eqref{eq.binary}.  This division yields a correlator equal to $\Enq=\frac14\times\frac12=\frac18$. 
Such feature of a system---that it can be split into no more than two mutually Bell-uncorrelated parts to reproduce the value of the correlator---we call 2-locality. 
{Note that the above value of the correlator could also be represented as $\Enq=\frac18=\frac12\times\frac12\times\frac12$. This is, however, possible only when the whole graph consits of three
particles, each yielding binary outcomes. In this work we consider larger graphs, for which, as stated above, $\Enq=\frac18$ signals 2-locallity.}

Similarly, when $\gamma=1$ and thus $l=2$, we obtain $\Enq=\frac1{16}$, which allows for a maximum of three uncorrelated parts: a GHZ state of $L-2$ qubits and two Bell-uncorrelated binary
particles [see again Eq.~\eqref{eq.binary}], yielding
$\Enq=\frac14\times\frac12\times\frac12=\frac1{16}$. Such state is 3-local. If the correlator is in the range $\Enq\!\in\,]\frac1{16},\frac18]$, which according to Eq.~\eqref{eq.depth.b} implies that
    $l=1$, then the 3-local partition is no longer possible and the state is maximally 2-local ($l+1=2$). This illustrates a general relation between the number $l$ from Eq.~\eqref{eq.depth.b},
    and $(l+1)$-locality.

The same applies to entanglement certification. When $\gamma$ satisfies
\begin{align}\label{eq.depth.en}
  s+1>\gamma\geqslant s
\end{align}
with $s\in\mathbb N$, then the state is maximally $(s+1)$-separable. For example, when $\Enq=\frac1{64}$, the state can be maximally decomposed into three groups,
      an $L-2$ GHZ state and two separable qubits, see Eqs~\eqref{eq.ghz.corr} and~\eqref{eq.bloch} respectively. This configuration
      yields $\Enq=\frac14\times\frac14\times\frac14$. When $\Enq\!\in\,]\frac1{64},\frac1{16}]$, so that $s=1$, this three-part decomposition is no longer possible and the state is maximally 2-separable.

We use the exponent $\gamma$ as a tool to study the quantum resources of graph states and demonstrate their potential for quantum metrology.

\subsection{Relation to quantum metrology}

The many-body correlations characterized by $\Enq$ are useful for quantum-enhanced metrology~\cite{PhysRevLett.126.210506}. 
According to the Cramer-Rao theorem, quantum Fisher information (QFI)~\cite{braunstein1994statistical} sets the lower bound for the sensitivity of estimating the phase $\theta$ through
\begin{align}
  \Delta^2\theta\geqslant{\cal F}^{-1}[\hat\varrho],
\end{align}
where $\hat\varrho$ is a density operator that carries information about an unknown parameter $\theta$. Given the spectral decomposition of the density matrix
\begin{align}
  \hat\varrho=\sum_ip_i\ketbra{\psi_i}{\psi_i}
\end{align}
the QFI is equal to 
\begin{align}
{\cal F}[\hat\varrho]=2\sum_{i,j}\frac1{p_i+p_j}\modsq{\bra{\psi_i}\dot{\hat\varrho}\ket{\psi_j}},
\end{align}
where the dot symbol denotes the derivative over $\theta$. If $\hat\varrho$ consists of $L$ non-entangled 
qubits and the transformation that introduces dependence of the density operator on $\theta$ is linear
(i.e., it consists of one-qubit operations), then
\begin{align}
  {\cal F}[\hat\varrho]\leqslant L,
\end{align}
known as the shot-noise limit~\cite{giovannetti2004quantum}. The maximum value of the QFI is 
\begin{align}\label{eq.hl}
  {\cal F}[\hat\varrho]=L^2,
\end{align}
known as the Heisenberg limit~\cite{pezze2009entanglement}. For our purposes, it is worth noting that the QFI can be lower-bounded using the correlator~\eqref{eq.elq} 
as follows~\cite{PhysRevLett.126.210506}
\begin{align}\label{eq.QFI}
  {\cal F}[\hat\varrho] \geqslant 4L^2 \Enq[\hat\varrho],
\end{align}
In particular, the maximum value of $\Enq=4^{-1}$, see Eqs~\eqref{eq.ghz} and~\eqref{eq.ghz.corr}, yields a sensitivity at the Heisenberg limit, see Eq.~\eqref{eq.hl}.
We also note that if
\begin{align}\label{eq.cond}
    \gamma < \log_4L,
\end{align}
then according to Eq.~\eqref{eq.QFI}, ${\cal F}[\hat\varrho]>L$, which yields the sub-shot noise sensitivity of phase estimation, i.e., $\Delta^2\theta<L^{-1}$.

\section{Results}

We demonstrate how the aforementioned methods can be applied to analyze various graph states.

\subsection{The star graph}

First, we start with a star graph and analyze how quantum correlations characterized by $\Enq$ depend on the number of connections between nodes. 
The initial setup consists of a central node connected to all the  other nodes but there are no boundaries connecting the corona
nodes, see Fig.~\ref{fig:fig1}(a). For $L$ qubits, this configuration corresponds to the GHZ state, and the optimal direction is given by $\vec{\kappa}^*=\{z_1,x_2\dots x_L\}$, i.e.
\begin{align}\label{eq.ghz.N}
  \ket{\psi}=\frac1{\sqrt2}\left(\ket{0^{(1)}_z0_x^{(2)}0_x^{(3)}\ldots}+\ket{1^{(1)}_z1^{(2)}_x1^{(3)}_x\ldots}\right).
\end{align}
Thus $\gamma=0$,  the entanglement and Bell correlations are fully $L$-partite,  and Eq.~\eqref{eq.ghz.N} is the reference state for quantum metrology.
We will now demonstrate that adding edges to this graph degrades the quantum correlations. The 
first edge connection
couples the second and third qubits via $\cz^{23}$ and transforms $\ket{0_x^{(2)}0_x^{(3)}}$ and $\ket{1_x^{(2)}1_x^{(3)}}$ of the state~\eqref{eq.ghz.N} into

\begin{subequations}\label{eq.action}
  \begin{align}
    &\cz^{23}\ket{0_x^{(2)}0_x^{(3)}}=\frac1{\sqrt2}(\ket{0_z^{(2)}0_x^{(3)}}+\ket{1_z^{(2)}1_x^{(3)}})\\
    &\cz^{23}\ket{1_x^{(2)}1_x^{(3)}}=\frac1{\sqrt2}(\ket{0_z^{(2)}1_x^{(3)}}-\ket{1_z^{(2)}0_x^{(3)}}).
  \end{align}
\end{subequations}
Therefore, the full graph state with one edge is
\begin{align}\label{eq.one.edge}
 \ket{\psi}=\frac{1}{2}\Big[\ket{0^{(1)}_z}\Big(\ket{0_z^{(2)}0_x^{(3)}}+\ket{1_z^{(2)}1_x^{(3)}}\Big)\ket{0_x^{(4)}...}+\ket{1^{(1)}_z}\Big(\ket{0_z^{(2)}1_x^{(3)}}-\ket{1_z^{(2)}0_x^{(3)}}\Big)\ket{1_x^{(4)}...}\Big].
\end{align}
Clearly, this state, expressed in the $x$- and $z$-bases, does not possess any GHZ component. However, transforming the states from Eqs~\eqref{eq.action} using the relation
\begin{subequations}                                                                                                    
  \begin{align}                                                                                                         
    \ket{0^{(k)}_x}&=\frac1{\sqrt2}(\ket{0^{(k)}_y}+i\ket{1^{(k)}_y})e^{-i\frac\pi4},\\
    \ket{1^{(k)}_x}&=\frac1{\sqrt2}(\ket{0^{(k)}_y}-i\ket{1^{(k)}_y})e^{i\frac\pi4}\\
    \ket{0^{(k)}_z}&=\frac1{\sqrt2}(\ket{0^{(k)}_y}+\ket{1^{(k)}_y}),\\
    \ket{1^{(k)}_z}&=\frac1{i\sqrt2}(\ket{0^{(k)}_y}-\ket{1^{(k)}_y}).                                                                          
  \end{align}                                                                                                           
\end{subequations}
gives 
\begin{subequations}\label{eq.23}
  \begin{align}
    &\cz^{23}\ket{0_x^{(2)}0_x^{(3)}}=\frac1{\sqrt2}(\ket{0_y^{(2)}0_y^{(3)}}+i\ket{1_y^{(2)}1_y^{(3)}})e^{-i\frac\pi4}\\
    &\cz^{23}\ket{1_x^{(2)}1_x^{(3)}}=\frac1{\sqrt2}(-\ket{0_y^{(2)}0_y^{(3)}}+i\ket{1_y^{(2)}1_y^{(3)}})e^{i\frac\pi4}.
  \end{align}
\end{subequations}
Inserting this expression into Eq.~\eqref{eq.one.edge} yields
\begin{align}\label{eq.star.two}
  \ket{\psi}=&\frac{1}{2}(\ket{0^{(1)}_z0_y^{(2)}0_y^{(3)}0_x^{(4)}...}-\ket{1^{(1)}_z1^{(2)}_y1^{(3)}_y1_x^{(4)}...}+\nonumber\\
  &i\ket{0^{(1)}_z1_y^{(2)}1_y^{(3)}0_x^{(4)}...}-i\ket{1^{(1)}_z0^{(2)}_y0^{(3)}_y1_x^{(4)}...})e^{-i\frac{\pi}4}.
\end{align}
Now, the GHZ terms can easily be recognized. The coupling of the first two terms is obtained with the  direction $\vec{\kappa}^*=\{z_1,y_2,y_3,x_4,\dots,x_L\}$, and the parameter $\gamma$ grows
from 0 (GHZ state) to $\gamma=1$. 
Therefore, locality or separability grows, compared with Eq.~\eqref{eq.ghz.N},
by either one or two parties, see Eqs~\eqref{eq.depth.b} and~\eqref{eq.depth.en} respectively. When analyzing subsequent couplings, we always identify the dominant GHZ component. 
By adding the connection between the third and fourth qubits via the $\cz^{(34)}$ gate and focusing only on the GHZ components from the first line of Eq.~\eqref{eq.star.two} the optimal direction becomes
\begin{align}
  \vec{\kappa}^*=\{z_1,y_2,y_3,x_4,\dots,x_L\} \to \vec{\kappa}^*=\{z_1,z_2,x_3,z_4,x_5,\dots,x_L\}
\end{align}
and the parameter grows to the value $\gamma=2$. 
Consecutive couplings  modify the optimal bases only for the pair of newly coupled nodes. For example, 
the next two couplings give the optimal  $\vec{\kappa}^*=\{z_1,z_2,x_3,z_4,z_5,x_6,\dots,x_L\}$, and $\gamma=3$. 
From this point on, a pattern emerges: whenever a link is added to an even number of links, $\gamma$ always increases by one. When added to an odd number of links, however, it remains the same. 
Interestingly, the closing connection
either keeps $\gamma$ constant (if $L$ is even) or increases it by 1 (if $L$ is odd). This general analysis allows us to predict the 
strength of quantum correlations in any star graph state~\cite{supp}.
\begin{figure}[t!]
    \centering
    \begin{overpic}[width= 0.8\linewidth]{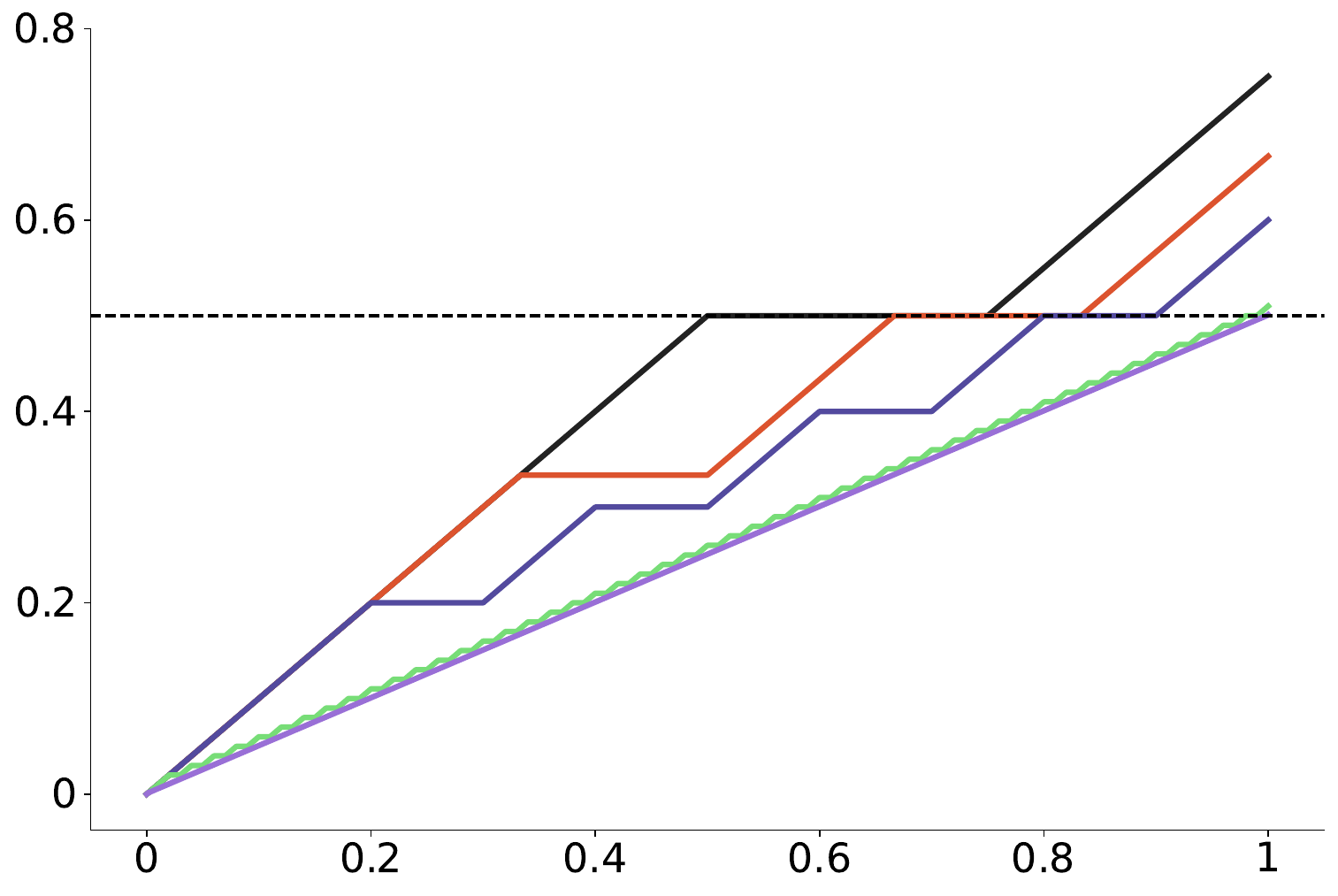}
    \put(55,-2){\makebox(0,0)[c]{normalized number of edges $x = p/L$}}
    \put(-5, 10){\rotatebox{90}{  normalized exponent $\gamma/L$}}
  \end{overpic}      
    \caption{The normalized exponent $\gamma/L$ for a star graph with edges, Fig.~\ref{fig:fig1}(b), is plotted as a function of
      the number of edges $p$ normalized to $L$, $x=p/L$. The colors correspond to:
      $L = 4$ (black), $L = 6$ (orange), $L=10^1$ (blue), $L=10^2$ (green) and $L=10^3$ (purple). The dashed horizontal line denotes the Bell limit. All the values of $\gamma/L<1$ 
     imply the presence of many-body entanglement, and the presence of many-body Bell correlations is indicated by values of $1/2 < \gamma/L < 1$.}
    \label{fig.n5}
\end{figure}

Figure~\ref{fig.n5} shows how the exponent $\gamma$, normalized to $L$, changes with an increasing number of established edges, using $L=4,6,10,100,1000$. 
The number of edges $p$ is normalized by $L$, yielding $x=p/L$, the fraction of edges.
For small $L$, the Bell correlations vanish (i.e., $\gamma/L>\frac12$), but in the quasi-thermodynamic limit (here we used $L=1000$ and checked that the curves do not change for higher $L$'s) 
the many-body correlations are present for  all star graph states.
Furthermore, for large $L$ we can analytically predict the strength of many-body Bell correlations and entanglement. Starting from $p=4$ edges, the $\gamma$
grows by one for every second $p$, hence the coherence behaves like $\gamma\simeq\frac p2$. According to Eq.~\eqref{eq.cond}, all graphs with $p<2\log_4L$ are useful for quantum metrology. 
One can also calculate the  strength of quantum correlations in the graph state.
For example, for $L=128$ and half of the nodes connected, i.e., $p=L/2$, we get $\gamma\simeq L/4$, hence according to Eq.~\eqref{eq.depth.b}, $l=64$ and the state is 
maximally $65$-local. Similarly, using Eq.~\eqref{eq.depth.en} we obtain $s=33$, so the sate is maximally 33-separable.

\subsection{The Tur\'an graph}
Next, we examine the topology of the Tur\'an graphs, and demonstrate their usefulness for quantum  metrology. The Tur\'an graph is defined
as follows: $L$ nodes are divided into $K$ groups, each with $n_k$ nodes, $\sum_{k=1}^K n_k = L$. 
Two nodes are connected if and only if they do not belong to the same group \cite{book_graphs}, see Fig.~\ref{fig:fig1}(c). 
A special case of a Tur\'an graph is a star graph with $K=2$ groups, where $n_1 = 1$, and $n_2 = L-1$. Considering a symmetric partition of nodes into $K$ groups the number of edges is
$|E| = (1-\frac{1}{K}) \frac{L^2-s^2}{2}+\binom{s}{2}$, where $s$ is a remainder of division $L/K$. For a fixed $L$, number of $\cz$ gates increases with the number of groups $K$.

Through the exact many-body calculations, we discovered that for graph states of $K$ groups, with at least two qubits in each group, $n_k\ge2$ for $k=1\dots K$, the exponent 
takes on the simple form $\gamma=K-1$ with the corresponding vector $\vec{\kappa}^*=\{x_1,\dots,x_L\}$. 
As such, $k$-locality and separability will increase with respect to the ideal GHZ state as the number of groups $K$ increases, 
see Eqs~\eqref{eq.depth.b} and ~\eqref{eq.depth.en}.
However, the fact that $\gamma$ depends only on $K$, and not on $L$, indicates that graph states with Tur\'an graph topology are valuable resources for quantum-enhanced metrology, 
see Eq.~\eqref{eq.cond}, i.e. $K<\log_4 L+1$ for sufficiently large $L$. We note that the {\it bundle graphs} introduced in Ref.\cite{PhysRevLett.124.110502} are in fact Tur\'an graphs. 

\subsection{The $r$-ary tree graph}\label{sec.trees}

In the following, we demonstrate how the method works for another prominent topology, the $r$-ary tree graphs~\cite{book_graphs}. These graphs are characterized  by the number
of links emanating from each node, $r$, and the distance between the center and the most distant node, $h$, see Fig.~\ref{fig:fig1}(d). 
Since the number of nodes in the $k$-th layer is $r^k$, the total number $L$ is
\begin{align}\label{eq.numb}
  L(r,h)=\sum_{k=0}^hr^k=\frac{1-r^{h+1}}{1-r}.
\end{align}
Analyzing quantum correlations in this system is most efficient when we first consider the star graphs in the shaded areas in  Fig.~\ref{fig:fig1}(d). 
Each star graph, disregarding its connections to other layers, forms a GHZ state
\begin{align}
  \ket{\psi_k}=\frac1{\sqrt2}(\ket{0_z}\ket{0_x}^{\otimes r^{k+1}}+\ket{1_z}\ket{1_x}^{\otimes r^{k+1}}).
\end{align}
There are $r^k$ GHZ states for the $k$\th  layer, formed at every second $k$, i.e., for $k=1,3,5\ldots$. 
Tree graphs with even and odd $r$ and $h$ should be considered separately because each coupling of the $\ket{1_z}$ component  
of the $(k+2)$\th GHZ state flips the $\ket{1_x}$'s of the $k$\th one. If $r$ is even ($e$), these flips return state to $\ket{1_x}$ while if it is odd ($o$), 
the result is $\ket{0_x}$. This drastically
reduces the GHZ coherence. For even $h$ the result is
\begin{align}
  \gamma_{e}=r\frac{r^h-1}{r^2-1},\ \ \ \gamma_{o}=\frac{r^h-2r+1}{r-1}.
\end{align}
Since the growth of $\gamma_{e,o}$ with $r$ and $h$ is smaller than the growth of the total number of nodes $L(r,h)$, Eq.\eqref{eq.numb}, 
we conclude that 
multi-branch trees are characterized by low $k$-locality/separability, see Eqs~\eqref{eq.depth.b} and~\eqref{eq.depth.en}. 
Adding another layer $h\rightarrow h+1$, that does not form a set of GHZ states [the nodes outside the red dashed circle in Fig.~\ref{fig:fig1}(d)], does not change the GHZ coherence of the tree graph. 
Therefore, the expressions for $\gamma_{e/o}$  hold. Crucially, when $r$ is odd $\gamma_o/L(r,h)\simeq r^{-2}$ which is small for large $r$~\cite{supp}. This implies strong quantum correlations, see
Eqs~\eqref{eq.depth.b} and~\eqref{eq.depth.en}.

\subsection{Cluster states}
Cluster states contain the resources necessary for measurement-based quantum computing~\cite{jozsa2006introduction,PhysRevLett.86.5188,PhysRevA.68.022312, doi:10.1142/S0219749904000055,
  PhysRevA.68.022312,hein2006entanglement,PhysRevLett.98.220503,briegel2009measurement,PhysRevLett.98.220503,Raza2023}. 
We considered cluster states on a $m\times n$ square lattice, see Fig.~\ref{fig:fig1}(e). Exact numerical calculations show
linear scaling of entanglement and Bell correlations with number of qubits $L=nm$, namely $\gamma = n+1$ for $m=2$, $\gamma = 2n-2$ for $m=3$.

\subsection{Non-equivalent classes of graph states under LU}

Two graph states with the same number of qubits $L$ are characterized by the same multipartite entanglement if they can be connected 
  by \emph{local unitaries} (LU)---arbitrary $\mathrm{SU}(2)$ gates applied to each qubit   independently, possibly followed by relabeling of qubits. 
  For stabilizer states there is a convenient restriction: for $L\le 8$ qubits, LU‐equivalence coincides with
  \emph{local Clifford} (LC) equivalence,
  so that one may focus on local Clifford gates  with no loss of generality.  In the graph picture, LC operations act through
  the purely combinatorial operation known as \emph{local complementation} (LC$_k$),
  performed around a chosen vertex $k$, together with ordinary graph isomorphisms ~\cite{BOUCHET199375,PhysRevA.69.022316,Adcock2020}.
  Therefore, the problem of classifying $L$-qubit graph states reduces to counting the orbits of simple graphs with $L$ vertices under the group generated by LC$_k$ moves and vertex relabelings. 
  In Refs.\cite{PhysRevA.69.062311,Cabello2009,PhysRevA.80.012102} such a classification was 
  carried out  up to $L=7$, and $L=8$ qubits. The long–standing LU$=$LC  conjecture for graph states was known to be \emph{true} for every graph on $L\le 8$ qubits and \emph{false}
  from $L=27$ upwards \cite{Ji2010}.  Recently, the LU-LC equivalence conjecture extended the “true” region out to $L=11$ \cite{Simon2025,Burchardt2025}, and later to $L=19$ qubits \cite{claudet2025}. 
  We note that a recently proposed graph state characterization with \emph{foliage partition}, 
  being a LC-invariant for graph states, allows a polynomial computational complexity ${\cal O}(L^3)$ in the number of qubits  \cite{Vandre2024Marginals,Burchardt2025_2}.
 
\begin{figure}[t!]
\includegraphics[width=1.\linewidth]{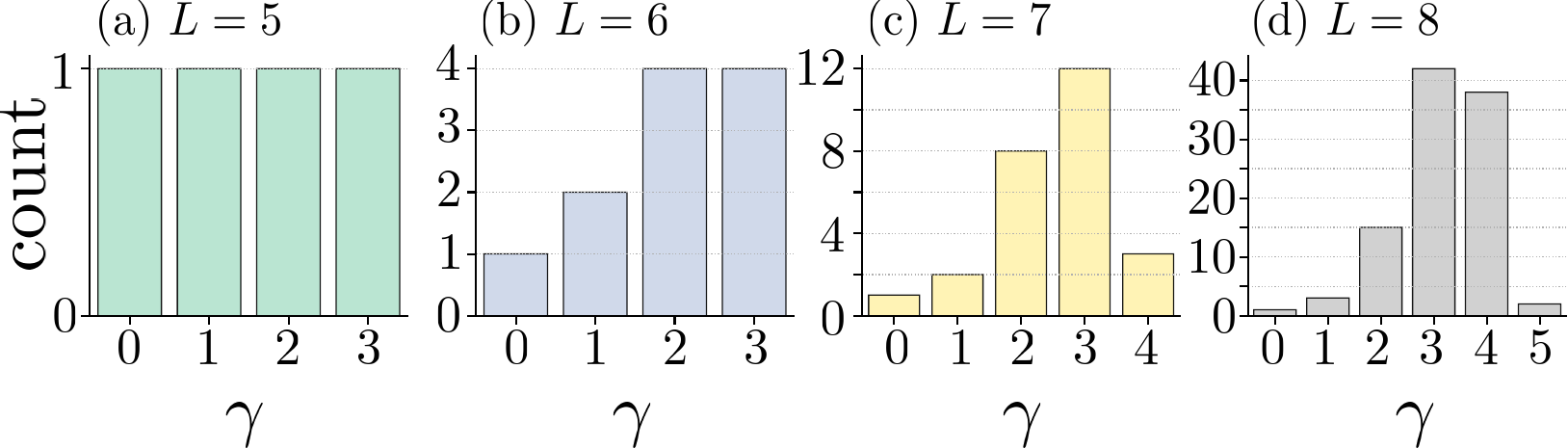} 
\caption{The histogram of $\gamma$ for LU non-equivalent graph states with $L=5,6,7,8$ qubits is shown in panels (a)-(d), respectively.}
    \label{fig_6}
\end{figure}

The catalog of $146$ LU (equivalently LC) classes is a complete   ``periodic table'' of graph-state entanglement with fewer than nine qubits. 
  For each class one can tabulate practical data such as the minimum number of controlled-$Z$ gates, the Schmidt ranks of all bi-partitions, and various
  multipartite entanglement measures. We characterize the structure of all $146$ non-equivalent LU classes
  for states  with  $L\leqslant 8$ qubits using $\gamma$, as shown in the figure at the following link~\cite{figure.lu}. In~Fig.\ref{fig_6} we show
  the histograms of $\gamma$ for $L = 5, 6, 7, 8$. These histograms show the most probable value of $\gamma$ for a graph state on a random connected graph of $L$-qubits.

\section{Conclusions}
We developed a method to characterize the quantum resources of graph states, including  many-body entanglement, Bell correlations, and quantum Fisher information. 
We provided analytical predictions for the many-body entanglement Bell correlations for any number of qubits for the considered examples of star, Tur'an, and r-ary tree graph topologies.
Furthermore, we demonstrated that Tur\'an graph topology constitutes a family of many-body quantum states that are useful for quantum metrology. We applied our  
method to analyze quantum correlations in square-lattice cluster states. 
Finally, we characterized  $146$ non-equivalent classes under local complementation and graph isomorphism for up to $8$ qubits with respect to $k$-separability.

The proposed method can be used for graph states with arbitrary topologies using a full many-body numerical approach. 
Alternatively, one can use a tensor network quantum state representation. Furthermore, the method allows us to
formulate the graph state preparation problem with a desired entanglement strength as an optimization problem with constraints on the qubit connectivity or the number of entangling gates.

The Bell correlator $\Enq$ allows to systematically trace the $k$-locality/separability of a quantum state. 
  As we have shown it is  sensitive to changes in the number of edges. Thus, one can observe how the exponent $\gamma$, which is roughly the distance
  of the state from the GHZ state, grows as the graph's connectivity increases.
  The correlator is given by the modulus squared of a single element of the density matrix. 
  Therefore,  it is relatively simple to calculate, even when one must optimize the orientation of the raising operator for each qubit. 
  Furthermore, the Bell correlator $\Enq$ should be accessible in experiments using the method of multiple quantum coherences~\cite{garttner2017measuring,PhysRevLett.120.040402}.

\section{Acknowledgments}JCh was supported by the National Science Centre, Poland, within the QuantERA II Programme that has received funding from the European Union’s Horizon 2020                      
research and innovation program under Grant Agreement No 101017733, Project No. 2021/03/Y/ST2/00195.

ICFO group acknowledges support from:
European Research Council AdG NOQIA; MCIN/AEI (PGC2018-0910.13039/501100011033, CEX2019-000910-S/10.13039/501100011033, Plan National FIDEUA PID2019-106901GB-I00, Plan National STAMEENA PID2022-139099NB, I00, project funded by MCIN/AEI/10.13039/ 501100011033 and by the “European Union NextGenerationEU/PRTR" (PRTR-C17.I1), FPI); 
QUANTERA MAQS PCI2019-111828-2; QUANTERA DYNAMITE PCI2022-132919, QuantERA II Programme co-funded by European Union’s Horizon 2020 program under Grant Agreement No 101017733; Ministry for Digital Transformation and of Civil Service of the Spanish Government through the QUANTUM ENIA project call - Quantum Spain project, and by the European Union through the Recovery, Transformation and Resilience Plan - NextGenerationEU within the framework of the Digital Spain 2026 Agenda; Fundació Cellex; Fundació Mir-Puig; Generalitat de Catalunya (European Social Fund FEDER and CERCA program, AGAUR Grant No. 2021 SGR 01452, QuantumCAT \ U16-011424, co-funded by ERDF Operational Program of Catalonia 2014-2020); Barcelona Supercomputing Center MareNostrum (FI-2023-3-0024);  Funded by the European Union. 

Views and opinions expressed are however those of the author(s) only and do not necessarily reflect those of the European Union, European Commission, European Climate, Infrastructure and Environment Executive Agency (CINEA), or any other granting authority.  Neither the European Union nor any granting authority can be held responsible for them (HORIZON-CL4-2022-QUANTUM-02-SGA  PASQuanS2.1, 101113690, EU Horizon 2020 FET-OPEN OPTOlogic, Grant No 899794),  EU Horizon Europe Program (This project has received funding from the European Union’s Horizon Europe research and innovation program under grant agreement No 101080086 NeQSTGrant Agreement 101080086 — NeQST);  ICFO Internal “QuantumGaudi” project;  European Union’s Horizon 2020 program under the Marie Sklodowska-Curie grant agreement No 847648; “La Caixa” Junior Leaders fellowships, La Caixa” Foundation (ID 100010434): CF/BQ/PR23/11980043.

\providecommand{\newblock}{}

\appendix

\title{Supplementary Material:\\Many-body quantum resources of  graph states}
\renewcommand{\theequation}{S.\arabic{equation}}


\section{Star graphs}\label{app.star}

For subsequent analysis it is important to list the following relations
\begin{subequations}\label{eq.cz.xx}
  \begin{align}
    \label{eq.11}&\hat C_z\ket{0_x0_x}=\frac1{\sqrt2}(\ket{0_z0_x}+\ket{1_z1_x})\\
    \label{eq.01}&\hat C_z\ket{1_x0_x}=\frac1{\sqrt2}(\ket{0_z0_x}-\ket{1_z1_x})\\
    \label{eq.10}&\hat C_z\ket{0_x1_x}=\frac1{\sqrt2}(\ket{0_z1_x}+\ket{1_z1_x})\\
    \label{eq.00}&\hat C_z\ket{1_x1_x}=\frac1{\sqrt2}(\ket{0_z1_x}-\ket{1_z0_x}).
  \end{align}
\end{subequations}
Therefore, connecting a pair of nodes in the eigenstate of the $x$ operator yields an additional factor of $1/\sqrt2$, which reduces the GHZ coherence. Conversely,
coupling one of node in the eigenstate of $x$ and the other in the eigenstate of $z$ does not produce such terms,
\begin{subequations}
  \begin{align}
    &\hat C_z\ket{0_z0_x}=\ket{0_z0_x}\\
    &\hat C_z\ket{0_z1_x}=\ket{0_z0_x}\\
    &\hat C_z\ket{1_z0_x}=\ket{1_z1_x}\\
    &\hat C_z\ket{1_z1_x}=\ket{1_z0_x}.
  \end{align}
\end{subequations}
A pair of nodes such as $\ket{0_z0_x}$ is an eigenstate of $\hat C_z$ with $\pm1$ eigenvalues. Therefore, they also do not influence GHZ coherence.

The analysis begins with the star state, which is the zero-border GHZ state
\begin{align}
  \ket{\psi}=\frac1{\sqrt2}\left(\ket{0^{(1)}_z0_x^{(2)}0_x^{(3)}\ldots}+\ket{1^{(1)}_z1^{(2)}_x1^{(3)}_x\ldots}\right),
\end{align}
with the optimal direction given by the vector $\vec{\kappa}^* = \{z_1,x_2,\dots ,x_L\}$.
The first connection via $\cz^{(23)}$ gate yields
\begin{align}
  \ket{\psi}&=\frac1{2}\Big(\ket{0^{(1)}_z}(\ket{0_y^{(2)}0_y^{(3)}}+i\ket{1_y^{(2)}1_y^{(3)}})e^{-i\frac\pi4}\ket{0_x^{(4)}\ldots}+\nonumber\\
  &-\ket{1^{(1)}_z}
  (\ket{0_y^{(2)}0_y^{(3)}}-i\ket{1_y^{(2)}1_y^{(3)}})e^{i\frac\pi4}\ket{1_x^{(4)}\ldots}\Big).
\end{align}
Note the $1/2$ prefactor, as opposed to the $1/\sqrt2$ of the GHZ state. This difference arises from the coupling of tho $x$-axis eigenstates, see Eqs~\eqref{eq.cz.xx}.
The dominant GHZ coherence is related to the states $\ket{0^{(1)}_z0_y^{(2)}0_y^{(3)}0_x^{(4)}\ldots}$
and $\ket{1^{(1)}_z1_y^{(2)}1_y^{(3)}1_x^{(4)}\ldots}$ with the prefactor $1/2$. Hence, the new optimal direction $\vec{\kappa}^*$ and the exponent $\gamma$ are equal to
\begin{align}
\vec{\kappa}^*&=\{z_1,y_2,y_3,x_4,\dots,x_L\}\\
  \gamma&=1
\end{align}
as reported in the main text.

The second border, i.e. $\cz^{(34)}$ gate, generates states that multiply the $\ket{0_z^{(1)}}$ and $\ket{1_z^{(1)}}$ of the central node as follows
\begin{subequations}
  \begin{align}
    &\ket{0_z^{(1)}}:\ \ \ \ket{0_z^{(2)}}\frac1{\sqrt2}(\ket{0^{(3)}_x0^{(4)}_z}+\ket{1^{(3)}_x1^{(4)}_z})+\ket{1_z^{(2)}}\frac1{\sqrt2}(\ket{1^{(3)}_x0^{(4)}_z}+\ket{0^{(3)}_x1^{(4)}_z})\\
    &\ket{1_z^{(1)}}:\ \ \ \ket{0_z^{(2)}}\frac1{\sqrt2}(\ket{0^{(3)}_x0^{(4)}_z}-\ket{1^{(3)}_x1^{(4)}_z})-\ket{1_z^{(2)}}\frac1{\sqrt2}(\ket{0^{(3)}_x0^{(4)}_z}-\ket{1^{(3)}_x1^{(4)}_z}).
  \end{align}
\end{subequations}
Once again, coupling the first term of the first line with the last term of the second line yields the maximum GHZ coherence, hence
\begin{align}
  \vec{\kappa}^*&=\{z_1,z_2,x_3,z_4,x_5,\dots,x_L\}\\
  \gamma&=2.
\end{align}
From now on, we will only analyze the evolution of these two terms, namely $\ket{0^{(1)}_z0^{(2)}_z0^{(3)}_x0^{(4)}_z}\ket{0_x}^{\otimes L-4}$ versus
$\ket{1_z^{(1)}1^{(2)}_z1^{(3)}_x1^{(4)}_z}\ket{1_x}^{\otimes L-4}$. Thus, for three connections the dominant GHZ terms are as follows:
\begin{subequations}
  \begin{align}
    &\ket{0^{(1)}_z0^{(2)}_z0^{(3)}_x0^{(4)}_z}\ket{0_x}^{\otimes L-4}\longrightarrow\ket{0^{(1)}_z0^{(2)}_z0^{(3)}_x0^{(4)}_z0^{(5)}_x}\ket{0_x}^{\otimes L-5}\\
    &\ket{1_z^{(1)}1^{(2)}_z1^{(3)}_x1^{(4)}_z}\ket{1_x}^{\otimes L-4}\longrightarrow\ket{1_z^{(1)}1^{(2)}_z1^{(3)}_x1^{(4)}_z0^{(5)}_x}\ket{1_x}^{\otimes L-5}.
  \end{align}
\end{subequations}
To extract the GHZ coherence, the 5\textsuperscript{th} node must be decomposed in a basis that is orthogonal to $x$, such as $z$. However, this produces an extra term $1/\sqrt2$ that decreases the
coherence, which is now equal to
\begin{align}
\vec{\kappa}^*&=\{z_1,z_2,x_3,z_4,z_5,x_6,\dots,x_L\}\\
  \gamma&=3.
\end{align}
Next, the 4\textsuperscript{th} connection yields:
\begin{subequations}
  \begin{align}
    &\ket{0^{(1)}_z0^{(2)}_z0^{(3)}_x0^{(4)}_z0^{(5)}_x}\ket{0_x}^{\otimes L-5}\longrightarrow\ket{0^{(1)}_z0^{(2)}_z0^{(3)}_x0^{(4)}_z0^{(5)}_x0^{(6)}_z}\ket{0_x}^{\otimes L-6}\\
    &\ket{1_z^{(1)}1^{(2)}_z1^{(3)}_x1^{(4)}_z0^{(5)}_x}\ket{1_x}^{\otimes L-5}\longrightarrow\ket{1_z^{(1)}1^{(2)}_z1^{(3)}_x1^{(4)}_z1^{(4)}_x1^{(5)}_z}\ket{1_x}^{\otimes L-6}.
  \end{align}
\end{subequations}
Now the extra rotation is unnecessary. Although the coupling generated another $1/\sqrt{2}$ it is offset by the lack of need to rotate the final node. This explains why the
GHZ coherence,
\begin{align}
  \vec{\kappa}^*&=\{z_1,z_2,x_3,z_4,x_5,z_6,x_7,\dots,x_L\}\\
  \gamma&=3
\end{align}
does not decrease compared to the three-connections case.
From now on the  pattern holds true: when $n$ is even
\begin{subequations}
  \begin{align}
    &\gamma_n=\gamma_{n-1},\ \ \ \vec{\kappa}^*=\{z_1,z_2,x_3,z_4,x_5,z_6,x_7,z_8,x_9,\ldots ,x_{L}\} \\
    &\gamma_{n+1}=1+\gamma_n,\ \ \ \vec{\kappa}^*=\{z_1,z_2,x_3,z_4,x_5,z_6,x_7,z_8,z_9,x_{10},\ldots ,x_L\}.
  \end{align}
\end{subequations}
The closing connection couples the $L$\th  with the 2\nd one. According to the above arguments, when $L$ is even this couples
$\ket{0_z^{(2)}}$ with $\ket{0_z^{(L)}}$ and $\ket{1_z^{(2)}}$ with $\ket{1_z^{(L)}}$. These are the two-node eigenstates of the $\hat C_z^{(2L)}$ gate. Therefore, the
the GHZ coherence does not change with respect to the previous step. However, if $L$ is odd, the coupling is
\begin{subequations}
  \begin{align}
    &\ket{0_z^{(2)}0_x^{(L)}}\longrightarrow\ket{0_z^{(2)}0_x^{(L)}}\\
    &\ket{1_z^{(2)}0_x^{(L)}}\longrightarrow\ket{1_z^{(2)}1_x^{(L)}}.
  \end{align}
\end{subequations}
hence, the coherence grows by a factor $4$, i.e., $\gamma$ drops by one, since no additional change of basis for the $L$\th node is required.
The optimal direction of spin-raising operators has the alternating structure, namely
\begin{align}
  \vec{\kappa}^*=\{z_1,z_2,x_3,z_4,x_5, \ldots, z_{L-1},x_L\}.
\end{align}

\section{Trees}\label{app.tree}

Here we provide the general formula for the correlator with any number of connectors $r$ and any distance $h$. However, this can be done by considering the parity of each parameter
separately.

\bigskip

$\bullet$ {\bf $h$ even, $r$ odd}

When $r$ is odd, the first layer of $r$ GHZ states' coupling to the central node yields
\begin{align}
  \ket\psi=\frac1{\sqrt2^r}(\ket{0^{(1)}_x}\ket{0_z}^{\otimes r}\ket{0_x}^{\otimes r^2}+\ket{1^{(1)}_x}\ket{1_z}^{\otimes r}\ket{1_x}^{\otimes r^2}).
\end{align}
Since $r$ is odd, coupling the next layer will flip all the $\ket{1_x}$ of the previous layer, resulting in the dominant term
\begin{align}
  \ket\psi=\frac1{\sqrt2^{r+r^3}}(\ket{0^{(1)}_x}\ket{0_z}^{\otimes(r+r^3)}\ket{0_x}^{\otimes r^4}+\ket{1^{(1)}_x}\ket{1_z}^{\otimes(r+r^3)}\ket{1_x}^{\otimes r^4}+\ldots)\ket{0_x}^{\otimes r^2}.
\end{align}
Since $h$ is even, the procedure ends with another complete layer. However, the last $\ket{1_x}$'s have no more layers to couple to so they flip into $\ket{0_x}$, hence the full state takes the form
(we list the dominant GHZ components)
\begin{align}\label{eq.hero}
  \ket\psi&=\frac1{\sqrt2^{r+r^3+\ldots+r^{h-1}}}
  (\ket{0^{(1)}_x}\ket{0_z}^{\otimes(r+r^3+\ldots r^{h-1})}\ket{0_x}^{\otimes r^h}+\nonumber\\
  &+\ket{1^{(1)}_x}\ket{1_z}^{\otimes(r+\ldots r^{h-1})}\ket{1_x}^{\otimes r^h})\ket{0_x}^{\otimes (r^2+\ldots+r^{h-2})}.
\end{align}
Therefore, the exponent $\gamma$ reads
\begin{align}
  \gamma=-1+r+r^2+r^3+\ldots+r^{h-1}
\end{align}

\bigskip

$\bullet$ {\bf $h$ even, $r$ even}

When $r$ is even, the first coupling yields
\begin{align}
  \ket\psi=\frac1{\sqrt2^r}\ket{0^{(1)}_x}(\ket{0_z}^{\otimes r}\ket{0_x}^{\otimes r^2}+\ket{1_z}^{\otimes r}\ket{1_x}^{\otimes r^2}).
\end{align}
Since $r$ is now even, the coupling of the next layer does not flip the $\ket{1_x}$'s of the previous layer. The dominant term now reads
\begin{align}
  \ket\psi=\frac1{\sqrt2^{r+r^3}}(\ket{0^{(1)}_x}\ket{0_z}^{\otimes(r+r^3)}\ket{0_x}^{\otimes (r^2+r^4)}+\ket{1^{(1)}_x}\ket{1_z}^{\otimes(r+r^3)}\ket{1_x}^{\otimes (r^2+r^4)}+\ldots).
\end{align}
Once again, as $h$ is even, the procedure concludes with an additional complete layer, resulting in the full state
\begin{align}
  \ket\psi=\frac1{\sqrt2^{r+r^3+\ldots+r^{h-1}}}
  \ket{0^{(1)}_x}(\ket{0_z}^{\otimes(r+r^3+\ldots r^{h-1})}\ket{0_x}^{\otimes r^2+\ldots r^h}+\ket{1_z}^{\otimes(r+\ldots r^{h-1})}\ket{1_x}^{\otimes r^2+\ldots+r^h}).
\end{align}
Therefore, the exponent $\gamma$ in this case reads
\begin{align}
  \gamma=-1+r+r^3+\ldots+r^{h-1}.
\end{align}

\bigskip

$\bullet$ {\bf $h$ even $\rightarrow$ $h+1$ odd, $r$ odd}

When an additional incomplete layer is added, its $\ket{0_x}$ states couple the to $\ket{0_x}$ states of the previous (complete layer). This transforms the latter
into $\ket{0_z}$ or $\ket{1_z}$ and generate another fractal of GHZ states. Thus, the state from Eq.~\eqref{eq.hero} transforms into
\begin{align}
  \ket\psi&=\frac1{\sqrt2^{r+r^3+\ldots+r^{h-1}+r^{h+1}}}
  (\ket{0^{(1)}_x}\ket{0_z}^{\otimes(r+r^3+\ldots r^{h-1}+r^h)}\ket{0_x}^{\otimes r^{h+1}}+\nonumber\\
  &+\ket{1^{(1)}_x}\ket{1_z}^{\otimes(r+\ldots r^{h-1}+r^{h})}\ket{1_x}^{\otimes r^{h_1}})\ket{0_x}^{\otimes (r^2+\ldots+r^{h-2}}.
\end{align}
Therefore,  the GHZ coherence in this case is
\begin{align}
  \gamma=-1+r+r^2+r^3+\ldots+r^{h-1}+r^{h+1}.
\end{align}

\bigskip

$\bullet$ {\bf $h$ even $\rightarrow$ $h+1$ odd, $r$ even}

Note that the same argument applies to $r$ being even, because these two cases do not differ with respect to the last complete layer. Therefore,
the state after adding the $h+1$-th incomplete layer is
\begin{align}
  \ket\psi&=\frac1{\sqrt2^{r+r^3+\ldots+r^{h-1}+r^{h+1}}}
  \ket{0^{(1)}_x}(\ket{0_z}^{\otimes(r+r^3+\ldots r^{h-1}+r^{h})}\ket{0_x}^{\otimes r^2+\ldots r^{h-2}+r^{h+1}}+\nonumber\\
  &+\ket{1_z}^{\otimes(r+\ldots r^{h-1}+r^h)}\ket{1_x}^{\otimes r^2+\ldots + r^{h-2}+r^{h+1}}).
\end{align}
Therefore, $\gamma$ in  this case reads
\begin{align}
  \gamma=-1 + r+r^3+\ldots+r^{h-1}+r^{h+1}.
\end{align}

\section{Non-equivalent classes under local unitaries and graph isomorphisms}

\begin{figure}[t!]\label{fig_sup_2}
    \includegraphics[width=0.99\linewidth]{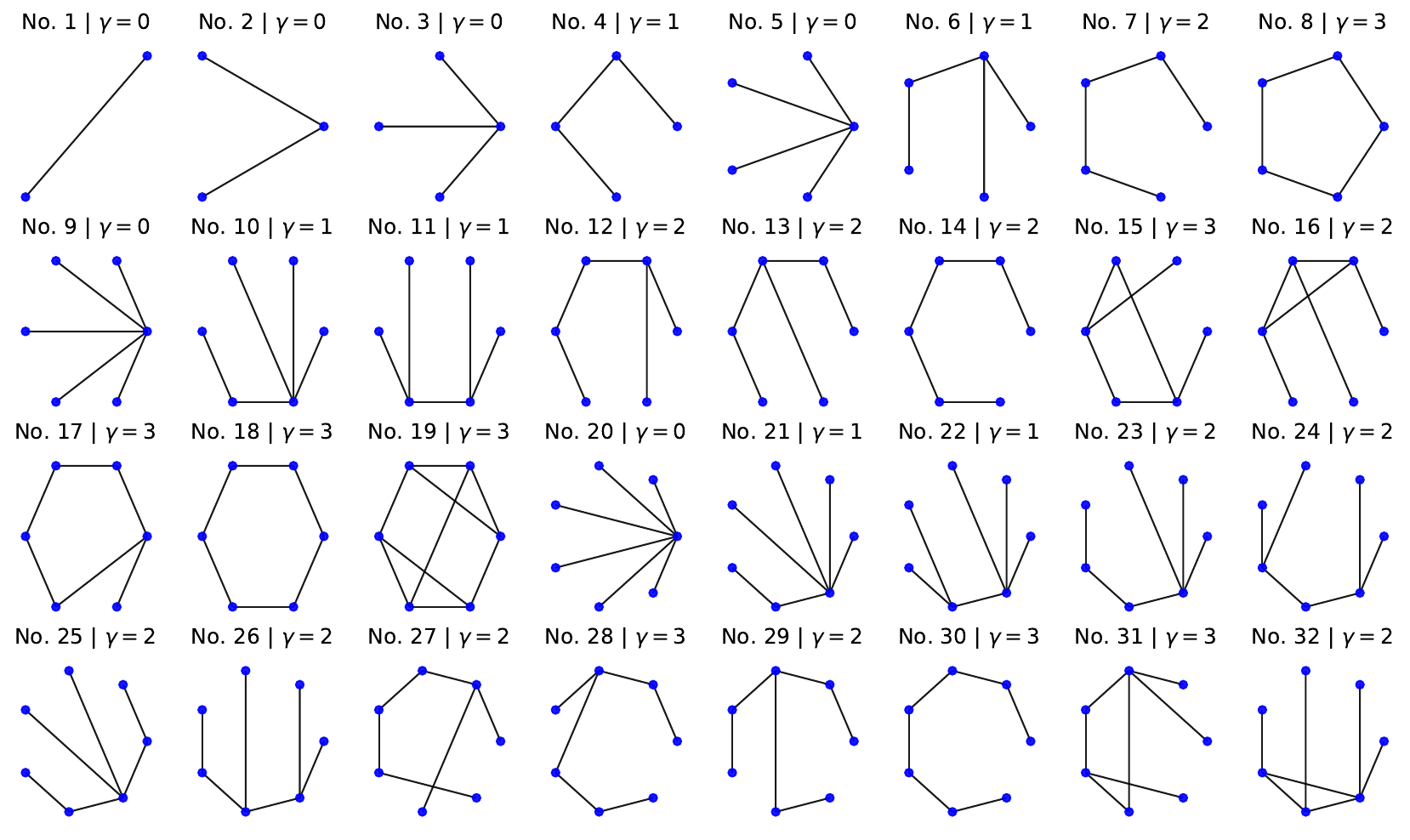}
    \includegraphics[width=0.99\linewidth]{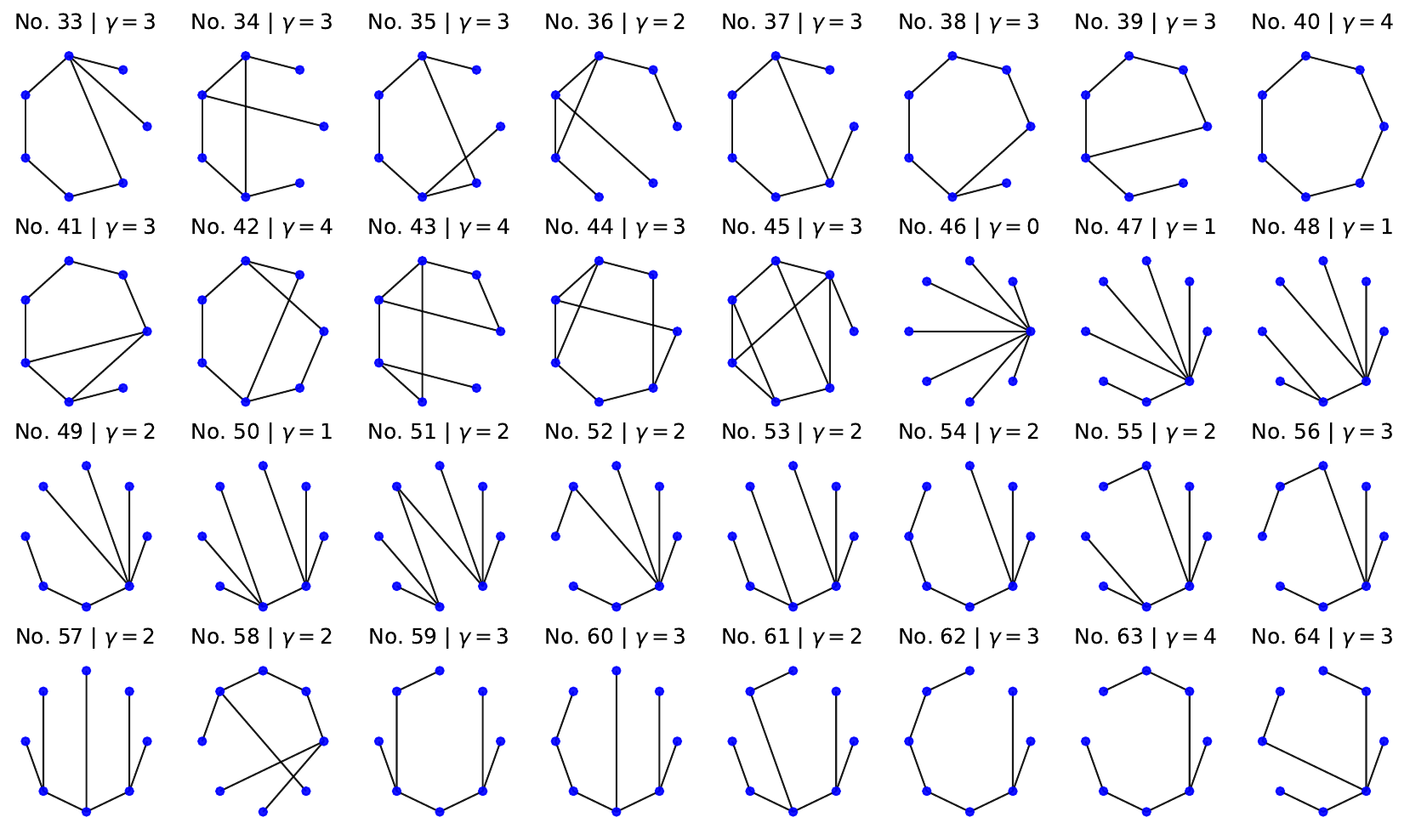}
    \caption{The exponent
$\gamma$ for representatives of graph states in non-equivalent classes under local unitary transformations and graph isomorphisms for up to
$L=8$ qubits. Classes No. 1-64.}
    \label{fig_sup_1}
\end{figure}

All connected graph states with up to $L = 8$ qubits can be classified into $146$ non-equivalent classes under local unitary transformations and graph isomorphisms \cite{PhysRevA.69.062311,Cabello2009,PhysRevA.80.012102}. Below, we present exponent $\gamma$ for each such a class representantive (graphs enumeration adapted from  \cite{PhysRevA.69.062311,Cabello2009,PhysRevA.80.012102}), see Fig.\ref{fig_sup_1}, Fig.\ref{fig_sup_2}, and Fig.\ref{fig_sup_3}.

\begin{figure}[t!]
\includegraphics[width=0.99\linewidth]{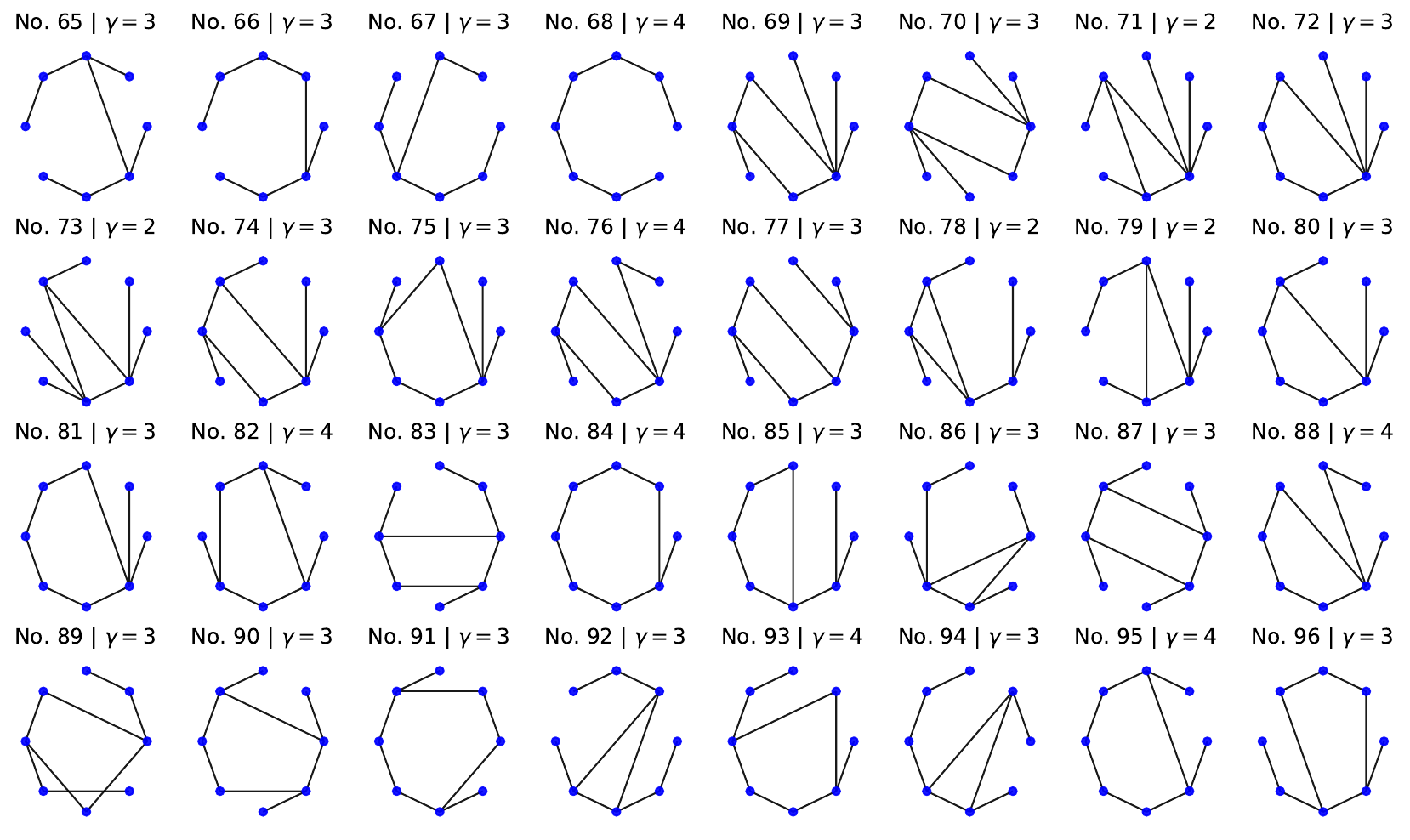}
\includegraphics[width=0.99\linewidth]{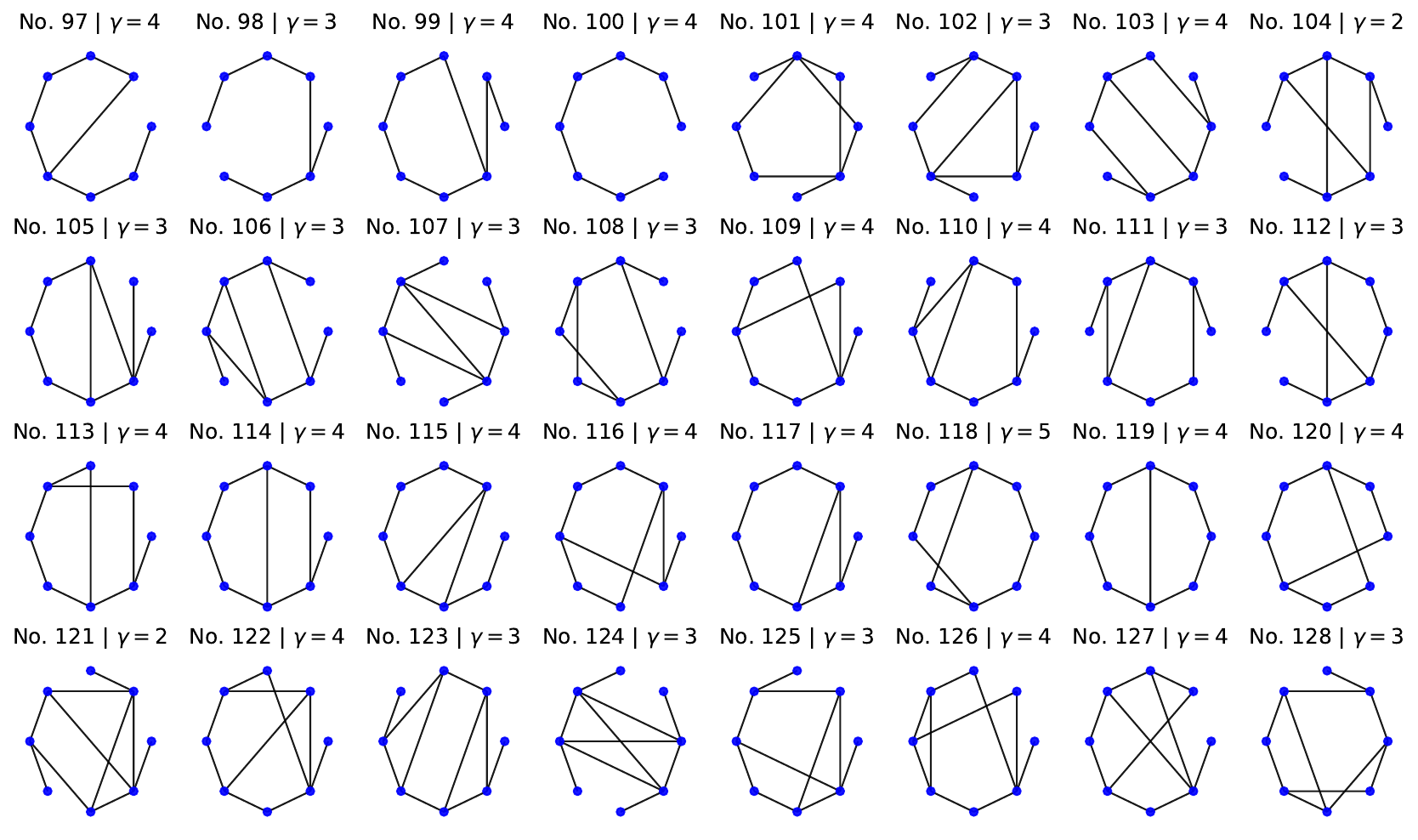}
    \caption{The exponent
$\gamma$ for representatives of graph states in non-equivalent classes under local unitary transformations and graph isomorphisms for up to
$L=8$ qubits. Classes No. 65-128.}
    \label{fig_sup_2}
\end{figure}

\begin{figure}[t!]
\includegraphics[width=0.99\linewidth]{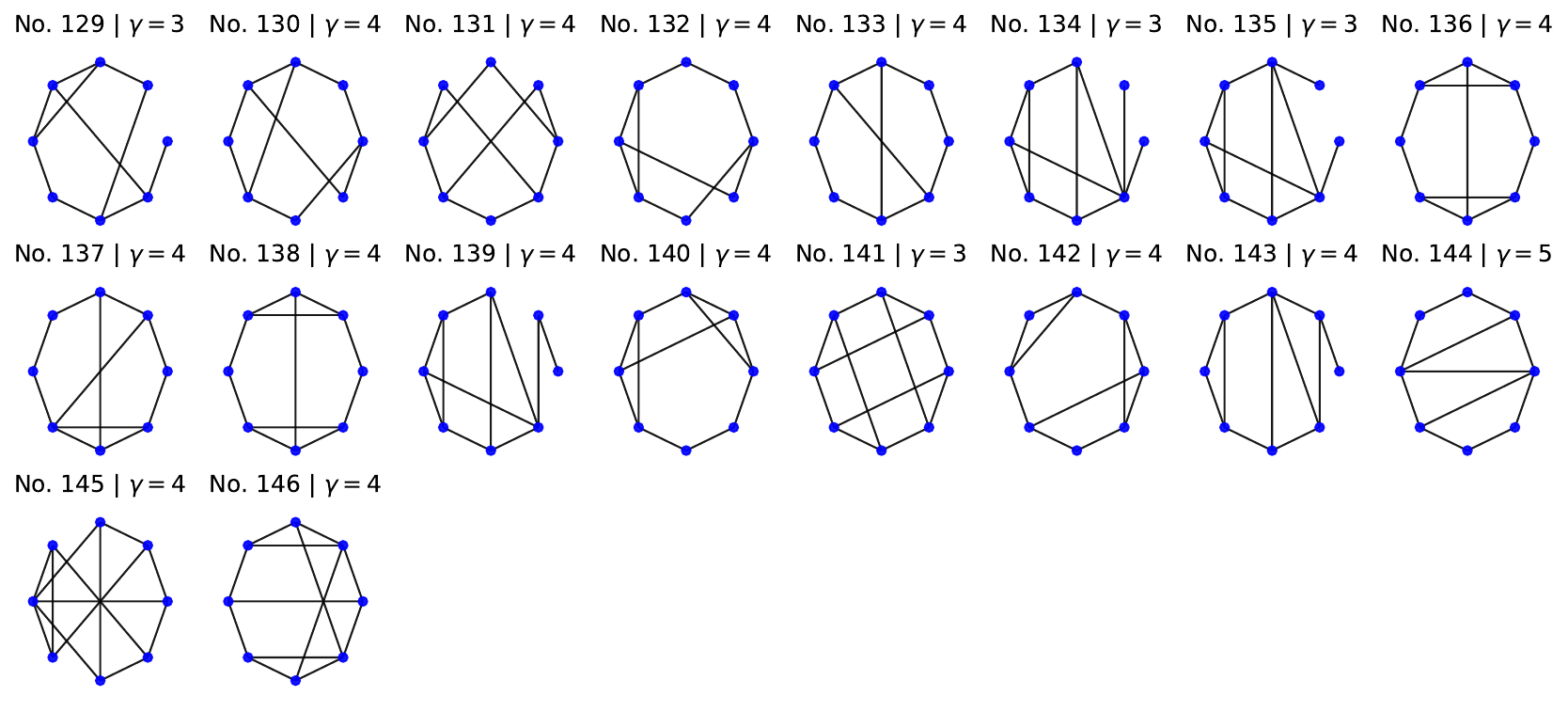}
    \caption{The exponent
$\gamma$ for representatives of graph states in non-equivalent classes under local unitary transformations and graph isomorphisms for up to
$L=8$ qubits. Classes No. 129-146.}
    \label{fig_sup_3}
\end{figure}


\begin{thebibliography}{10}
\expandafter\ifx\csname url\endcsname\relax
  \def\url#1{{\tt #1}}\fi
\expandafter\ifx\csname urlprefix\endcsname\relax\def\urlprefix{URL }\fi
\providecommand{\eprint}[2][]{\url{#2}}

\bibitem{RevModPhys.89.041003}
Streltsov A, Adesso G and Plenio M~B 2017 {\em Rev. Mod. Phys.\/} {\bf 89}(4)
  041003 \urlprefix\url{https://link.aps.org/doi/10.1103/RevModPhys.89.041003}

\bibitem{horodecki2009entanglement}
Horodecki R, Horodecki P, Horodecki M and Horodecki K 2009 {\em Rev. Mod.
  Phys.\/} {\bf 81}(2) 865--942
  \urlprefix\url{https://link.aps.org/doi/10.1103/RevModPhys.81.865}

\bibitem{brunner2014bell}
Brunner N, Cavalcanti D, Pironio S, Scarani V and Wehner S 2014 {\em Rev. Mod.
  Phys.\/} {\bf 86}(2) 419--478
  \urlprefix\url{https://link.aps.org/doi/10.1103/RevModPhys.86.419}

\bibitem{Srivastava2024}
Srivastava A~K, M\"{u}ller-Rigat G, Lewenstein M and Rajchel-Mieldzio\'c G 2024
  {\em Introduction to Quantum Entanglement in Many-Body Systems\/} (Springer
  Nature Switzerland) p 225–285 ISBN 9783031556579
  \urlprefix\url{http://dx.doi.org/10.1007/978-3-031-55657-9-4}

\bibitem{horodecki2024multipartiteentanglement}
Horodecki P, Łukasz Rudnicki and Życzkowski K 2024 Multipartite entanglement
  (\textit{Preprint} \eprint{2409.04566})
  \urlprefix\url{https://arxiv.org/abs/2409.04566}

\bibitem{RevModPhys.91.025001}
Chitambar E and Gour G 2019 {\em Rev. Mod. Phys.\/} {\bf 91}(2) 025001
  \urlprefix\url{https://link.aps.org/doi/10.1103/RevModPhys.91.025001}

\bibitem{https://doi.org/10.48550/arxiv.2405.05785}
Salazar R, Czartowski J, Rodríguez R~R, Rajchel-Mieldzioć G, Horodecki P and
  Życzkowski K 2024 Quantum resource theories beyond convexity
  \urlprefix\url{https://arxiv.org/abs/2405.05785}

\bibitem{jozsa2006introduction}
Jozsa R 2006 {\em NATO Science Series, III: Computer and Systems Sciences.
  Quantum Information Processing-From Theory to Experiment\/} {\bf 199}
  137--158

\bibitem{PhysRevLett.86.5188}
Raussendorf R and Briegel H~J 2001 {\em Phys. Rev. Lett.\/} {\bf 86}(22)
  5188--5191
  \urlprefix\url{https://link.aps.org/doi/10.1103/PhysRevLett.86.5188}

\bibitem{PhysRevA.68.022312}
Raussendorf R, Browne D~E and Briegel H~J 2003 {\em Phys. Rev. A\/} {\bf 68}(2)
  022312 \urlprefix\url{https://link.aps.org/doi/10.1103/PhysRevA.68.022312}

\bibitem{doi:10.1142/S0219749904000055}
Leung D~W 2004 {\em International Journal of Quantum Information\/} {\bf 02}
  33--43 (\textit{Preprint} \eprint{https://doi.org/10.1142/S0219749904000055})
  \urlprefix\url{https://doi.org/10.1142/S0219749904000055}

\bibitem{hein2006entanglement}
Hein M, Dür W, Eisert J, Raussendorf R, den Nest M~V and Briegel H~J 2006
  Entanglement in graph states and its applications (\textit{Preprint}
  \eprint{quant-ph/0602096})

\bibitem{PhysRevLett.98.220503}
Gross D and Eisert J 2007 {\em Phys. Rev. Lett.\/} {\bf 98}(22) 220503
  \urlprefix\url{https://link.aps.org/doi/10.1103/PhysRevLett.98.220503}

\bibitem{briegel2009measurement}
Briegel H~J, Browne D~E, D{\"u}r W, Raussendorf R and Van~den Nest M 2009 {\em
  Nature Physics\/} {\bf 5} 19--26

\bibitem{Raza2023}
Raza M~A, Alahmadi A~N, Basaffar W, Glynn D~G, Gupta M~K, Hirschfeld J~W~P,
  Khan A~N, Shoaib H and Solé P 2023 {\em Mathematics\/} {\bf 11} 2310 ISSN
  2227-7390 \urlprefix\url{http://dx.doi.org/10.3390/math11102310}

\bibitem{PhysRevA.86.042304}
Cuquet M and Calsamiglia J 2012 {\em Phys. Rev. A\/} {\bf 86}(4) 042304
  \urlprefix\url{https://link.aps.org/doi/10.1103/PhysRevA.86.042304}

\bibitem{Bell2014}
Bell B~A, Markham D, Herrera-Martí D~A, Marin A, Wadsworth W~J, Rarity J~G and
  Tame M~S 2014 {\em Nature Communications\/} {\bf 5} ISSN 2041-1723
  \urlprefix\url{http://dx.doi.org/10.1038/ncomms6480}

\bibitem{PhysRevA.100.052333}
Meignant C, Markham D and Grosshans F 2019 {\em Phys. Rev. A\/} {\bf 100}(5)
  052333 \urlprefix\url{https://link.aps.org/doi/10.1103/PhysRevA.100.052333}

\bibitem{Javelle2013}
Javelle J, Mhalla M and Perdrix S 2013 {\em New Protocols and Lower Bounds for
  Quantum Secret Sharing with Graph States\/} (Springer Berlin Heidelberg) p
  1–12 ISBN 9783642356568
  \urlprefix\url{http://dx.doi.org/10.1007/978-3-642-35656-8-1}

\bibitem{Hahn2019}
Hahn F, Pappa A and Eisert J 2019 {\em npj Quantum Information\/} {\bf 5} ISSN
  2056-6387 \urlprefix\url{http://dx.doi.org/10.1038/s41534-019-0191-6}

\bibitem{Hilaire2021}
Hilaire P, Barnes E and Economou S~E 2021 {\em Quantum\/} {\bf 5} 397 ISSN
  2521-327X \urlprefix\url{http://dx.doi.org/10.22331/q-2021-02-15-397}

\bibitem{Ding2023}
Ding Y, Wei Y, Li Z and Jiang M 2023 {\em Quantum Information Processing\/}
  {\bf 22} ISSN 1573-1332
  \urlprefix\url{http://dx.doi.org/10.1007/s11128-023-04157-0}

\bibitem{Fischer2021}
Fischer A and Towsley D 2021 Distributing graph states across quantum networks
  {\em 2021 IEEE International Conference on Quantum Computing and Engineering
  (QCE)\/} (IEEE) p 324–333
  \urlprefix\url{http://dx.doi.org/10.1109/QCE52317.2021.00049}

\bibitem{Unnikrishnan2022}
Unnikrishnan A and Markham D 2022 {\em Physical Review A\/} {\bf 105} ISSN
  2469-9934 \urlprefix\url{http://dx.doi.org/10.1103/PhysRevA.105.052420}

\bibitem{PhysRevA.108.012402}
Meyer U~I, Grosshans F and Markham D 2023 {\em Phys. Rev. A\/} {\bf 108}(1)
  012402 \urlprefix\url{https://link.aps.org/doi/10.1103/PhysRevA.108.012402}

\bibitem{https://doi.org/10.48550/arxiv.2407.01429}
Li B, Goodenough K, Rozp\'edek F and Jiang L 2024 Generalized quantum repeater
  graph states \urlprefix\url{https://arxiv.org/abs/2407.01429}

\bibitem{Makuta2023}
Makuta O, Ligthart L~T and Augusiak R 2023 {\em npj Quantum Information\/} {\bf
  9} ISSN 2056-6387
  \urlprefix\url{http://dx.doi.org/10.1038/s41534-023-00789-3}

\bibitem{li2024}
Li B, Goodenough K, Rozp\'edek F and Jiang L 2024 Generalized quantum repeater
  graph states (\textit{Preprint} \eprint{2407.01429})
  \urlprefix\url{https://arxiv.org/abs/2407.01429}

\bibitem{PhysRevLett.124.110502}
Shettell N and Markham D 2020 {\em Phys. Rev. Lett.\/} {\bf 124}(11) 110502
  \urlprefix\url{https://link.aps.org/doi/10.1103/PhysRevLett.124.110502}

\bibitem{BOUCHET199375}
Bouchet A 1993 {\em Discrete Mathematics\/} {\bf 114} 75--86 ISSN 0012-365X
  \urlprefix\url{https://www.sciencedirect.com/science/article/pii/0012365X9390357Y}

\bibitem{PhysRevA.69.022316}
Van~den Nest M, Dehaene J and De~Moor B 2004 {\em Phys. Rev. A\/} {\bf 69}(2)
  022316 \urlprefix\url{https://link.aps.org/doi/10.1103/PhysRevA.69.022316}

\bibitem{Adcock2020}
Adcock J~C, Morley-Short S, Dahlberg A and Silverstone J~W 2020 {\em Quantum\/}
  {\bf 4} 305 ISSN 2521-327X
  \urlprefix\url{http://dx.doi.org/10.22331/q-2020-08-07-305}

\bibitem{PhysRevA.83.042314}
Cabello A, Danielsen L~E, L\'opez-Tarrida A~J and Portillo J~R 2011 {\em Phys.
  Rev. A\/} {\bf 83}(4) 042314
  \urlprefix\url{https://link.aps.org/doi/10.1103/PhysRevA.83.042314}

\bibitem{Russo2019}
Russo A, Barnes E and Economou S~E 2019 {\em New Journal of Physics\/} {\bf 21}
  055002 ISSN 1367-2630
  \urlprefix\url{http://dx.doi.org/10.1088/1367-2630/ab193d}

\bibitem{Thomas2024}
Thomas P, Ruscio L, Morin O and Rempe G 2024 {\em Nature\/} {\bf 629} 567–572
  ISSN 1476-4687 \urlprefix\url{http://dx.doi.org/10.1038/s41586-024-07357-5}

\bibitem{PRXQuantum.5.020346}
Raissi Z, Barnes E and Economou S~E 2024 {\em PRX Quantum\/} {\bf 5}(2) 020346
  \urlprefix\url{https://link.aps.org/doi/10.1103/PRXQuantum.5.020346}

\bibitem{PhysRevA.109.042604}
Li X~X, Li D~X and Shao X~Q 2024 {\em Phys. Rev. A\/} {\bf 109}(4) 042604
  \urlprefix\url{https://link.aps.org/doi/10.1103/PhysRevA.109.042604}

\bibitem{Clark2005}
Clark S~R, Alves C~M and Jaksch D 2005 {\em New Journal of Physics\/} {\bf 7}
  124–124 ISSN 1367-2630
  \urlprefix\url{http://dx.doi.org/10.1088/1367-2630/7/1/124}

\bibitem{PhysRevA.83.010302}
Ballester D, Cho J and Kim M~S 2011 {\em Phys. Rev. A\/} {\bf 83}(1) 010302
  \urlprefix\url{https://link.aps.org/doi/10.1103/PhysRevA.83.010302}

\bibitem{Cooper2024}
Cooper E~S, Kunkel P, Periwal A and Schleier-Smith M 2024 {\em Nature
  Physics\/} {\bf 20} 770–775 ISSN 1745-2481
  \urlprefix\url{http://dx.doi.org/10.1038/s41567-024-02407-1}

\bibitem{Li2020}
Li J~P, Qin J, Chen A, Duan Z~C, Yu Y, Huo Y, H\"{o}fling S, Lu C~Y, Chen K and
  Pan J~W 2020 {\em ACS Photonics\/} {\bf 7} 1603–1610 ISSN 2330-4022
  \urlprefix\url{http://dx.doi.org/10.1021/acsphotonics.0c00192}

\bibitem{Gnatenko2021}
Gnatenko K and Tkachuk V 2021 {\em Physics Letters A\/} {\bf 396} 127248 ISSN
  0375-9601 \urlprefix\url{http://dx.doi.org/10.1016/j.physleta.2021.127248}

\bibitem{Markham2007}
Markham D, Miyake A and Virmani S 2007 {\em New Journal of Physics\/} {\bf 9}
  194–194 ISSN 1367-2630
  \urlprefix\url{http://dx.doi.org/10.1088/1367-2630/9/6/194}

\bibitem{PhysRevA.64.022306}
Eisert J and Briegel H~J 2001 {\em Phys. Rev. A\/} {\bf 64}(2) 022306
  \urlprefix\url{https://link.aps.org/doi/10.1103/PhysRevA.64.022306}

\bibitem{Guo_2014}
Guo Q~Q, Chen X~Y and Wang Y~Y 2014 {\em Chinese Physics B\/} {\bf 23} 050309
  ISSN 1674-1056
  \urlprefix\url{http://dx.doi.org/10.1088/1674-1056/23/5/050309}

\bibitem{10.1063/1.4903126}
Hajdušek M and Murao M 2014 {\em AIP Conference Proceedings\/} {\bf 1633}
  168--170 ISSN 0094-243X (\textit{Preprint}
  \eprint{https://pubs.aip.org/aip/acp/article-pdf/1633/1/168/12136622/168\_1\_online.pdf})
  \urlprefix\url{https://doi.org/10.1063/1.4903126}

\bibitem{PhysRevLett.94.060501}
T\'oth G and G\"uhne O 2005 {\em Phys. Rev. Lett.\/} {\bf 94}(6) 060501
  \urlprefix\url{https://link.aps.org/doi/10.1103/PhysRevLett.94.060501}

\bibitem{zander2024}
Zander R and Becker C~K~U 2024 Benchmarking multipartite entanglement
  generation with graph states (\textit{Preprint} \eprint{2402.00766})
  \urlprefix\url{https://arxiv.org/abs/2402.00766}

\bibitem{zukowski2002bell}
{\.Z}ukowski M and Brukner {\v{C}} 2002 {\em Phys. Rev. Lett.\/} {\bf 88}
  210401

\bibitem{mermin1990extreme}
Mermin N~D 1990 {\em Phys. Rev. Lett.\/} {\bf 65} 1838

\bibitem{10.1119/1.12594}
Mermin N~D 1981 {\em American Journal of Physics\/} {\bf 49} 940--943 ISSN
  0002-9505 (\textit{Preprint}
  \eprint{https://pubs.aip.org/aapt/ajp/article-pdf/49/10/940/11864850/940\_1\_online.pdf})
  \urlprefix\url{https://doi.org/10.1119/1.12594}

\bibitem{cavalcanti2007bell}
Cavalcanti E~G, Foster C~J, Reid M~D and Drummond P~D 2007 {\em Phys. Rev.
  Lett.\/} {\bf 99} 210405

\bibitem{Reid2012}
Reid M~D, He Q~Y and Drummond P~D 2012 {\em Frontiers of Physics\/} {\bf 7}
  72--85 ISSN 2095-0470
  \urlprefix\url{https://doi.org/10.1007/s11467-011-0233-9}

\bibitem{cavalcanti2011unified}
Cavalcanti E, He Q, Reid M and Wiseman H 2011 {\em Phys. Rev. A\/} {\bf 84}
  032115

\bibitem{Ardehali_1992}
Ardehali M 1992 {\em Physical Review A\/} {\bf 46} 5375--5378 ISSN 1094-1622
  \urlprefix\url{http://dx.doi.org/10.1103/PhysRevA.46.5375}

\bibitem{garttner2017measuring}
G{\"a}rttner M, Bohnet J~G, Safavi-Naini A, Wall M~L, Bollinger J~J and Rey A~M
  2017 {\em Nature Physics\/} {\bf 13} 781--786

\bibitem{PhysRevLett.120.040402}
G\"arttner M, Hauke P and Rey A~M 2018 {\em Phys. Rev. Lett.\/} {\bf 120}(4)
  040402
  \urlprefix\url{https://link.aps.org/doi/10.1103/PhysRevLett.120.040402}

\bibitem{book_graphs}
Bondy J and Muny U 1976 {\em Graph theory with applications\/} (London: The
  Macmillan Press)

\bibitem{spiny.milosz}
Niezgoda A, Panfil M and Chwede\ifmmode~\acute{n}\else \'{n}\fi{}czuk J 2020
  {\em Phys. Rev. A\/} {\bf 102}(4) 042206
  \urlprefix\url{https://link.aps.org/doi/10.1103/PhysRevA.102.042206}

\bibitem{PhysRevA.102.013328}
P\l{}odzie\ifmmode~\acute{n}\else \'{n}\fi{} M, Ko\ifmmode~\acute{s}\else
  \'{s}\fi{}cielski M, Witkowska E and Sinatra A 2020 {\em Phys. Rev. A\/} {\bf
  102}(1) 013328
  \urlprefix\url{https://link.aps.org/doi/10.1103/PhysRevA.102.013328}

\bibitem{PhysRevLett.126.210506}
Niezgoda A and Chwede\ifmmode~\acute{n}\else \'{n}\fi{}czuk J 2021 {\em Phys.
  Rev. Lett.\/} {\bf 126}(21) 210506
  \urlprefix\url{https://link.aps.org/doi/10.1103/PhysRevLett.126.210506}

\bibitem{PhysRevLett.129.250402}
P\l{}odzie\ifmmode~\acute{n}\else \'{n}\fi{} M, Lewenstein M, Witkowska E and
  Chwede\ifmmode~\acute{n}\else \'{n}\fi{}czuk J 2022 {\em Phys. Rev. Lett.\/}
  {\bf 129}(25) 250402
  \urlprefix\url{https://link.aps.org/doi/10.1103/PhysRevLett.129.250402}

\bibitem{PhysRevA.107.013311}
Dziurawiec M, Hern\'andez~Yanes T, P\l{}odzie\ifmmode~\acute{n}\else \'{n}\fi{}
  M, Gajda M, Lewenstein M and Witkowska E 2023 {\em Phys. Rev. A\/} {\bf
  107}(1) 013311
  \urlprefix\url{https://link.aps.org/doi/10.1103/PhysRevA.107.013311}

\bibitem{PhysRevB.108.104301}
Hern\'andez~Yanes T, \ifmmode~\check{Z}\else \v{Z}\fi{}labys G,
  P\l{}odzie\ifmmode~\acute{n}\else \'{n}\fi{} M, Burba D, Sinkevi\ifmmode
  \check{c}\else \v{c}\fi{}ien\ifmmode~\dot{e}\else \.{e}\fi{} M~M, Witkowska E
  and Juzeli\ifmmode~\bar{u}\else \={u}\fi{}nas G 2023 {\em Phys. Rev. B\/}
  {\bf 108}(10) 104301
  \urlprefix\url{https://link.aps.org/doi/10.1103/PhysRevB.108.104301}

\bibitem{PhysRevResearch.6.023050}
P\l{}odzie\ifmmode~\acute{n}\else \'{n}\fi{} M, Wasak T, Witkowska E,
  Lewenstein M and Chwede\ifmmode~\acute{n}\else \'{n}\fi{}czuk J 2024 {\em
  Phys. Rev. Res.\/} {\bf 6}(2) 023050
  \urlprefix\url{https://link.aps.org/doi/10.1103/PhysRevResearch.6.023050}

\bibitem{PhysRevA.110.012210}
Hamza D~A and Chwede\ifmmode~\acute{n}\else \'{n}\fi{}czuk J 2024 {\em Phys.
  Rev. A\/} {\bf 110}(1) 012210
  \urlprefix\url{https://link.aps.org/doi/10.1103/PhysRevA.110.012210}

\bibitem{grze_mar}
P\l{}odzie\ifmmode~\acute{n}\else \'{n}\fi{} M, Chwede\ifmmode~\acute{n}\else
  \'{n}\fi{}czuk J, Lewenstein M and Rajchel-Mieldzio\ifmmode~\acute{c}\else
  \'{c}\fi{} G 2024 {\em Phys. Rev. A\/} {\bf 110}(3) 032428
  \urlprefix\url{https://link.aps.org/doi/10.1103/PhysRevA.110.032428}

\bibitem{plodzien2024inherent}
P\l{}odzie\ifmmode~\acute{n}\else \'{n}\fi{} M, Chwede\ifmmode~\acute{n}\else
  \'{n}\fi{}czuk J and Lewenstein M 2025 {\em Phys. Rev. A\/} {\bf 111}(1)
  012417 \urlprefix\url{https://link.aps.org/doi/10.1103/PhysRevA.111.012417}

\bibitem{braunstein1994statistical}
Braunstein S~L and Caves C~M 1994 {\em Phys. Rev. Lett.\/} {\bf 72} 3439--3443

\bibitem{giovannetti2004quantum}
Giovannetti V, Lloyd S and Maccone L 2004 {\em Science\/} {\bf 306} 1330--1336

\bibitem{pezze2009entanglement}
Pezz{\'e} L and Smerzi A 2009 {\em Phys. Rev. Lett.\/} {\bf 102} 100401

\bibitem{supp}
 {\em See the Supplementary materials for the detailed step-by-step derivation
  and discussion\/}

\bibitem{PhysRevA.69.062311}
Hein M, Eisert J and Briegel H~J 2004 {\em Phys. Rev. A\/} {\bf 69}(6) 062311
  \urlprefix\url{https://link.aps.org/doi/10.1103/PhysRevA.69.062311}

\bibitem{Cabello2009}
Cabello A, López-Tarrida A~J, Moreno P and Portillo J~R 2009 {\em Physics
  Letters A\/} {\bf 373} 2219–2225 ISSN 0375-9601
  \urlprefix\url{http://dx.doi.org/10.1016/j.physleta.2009.04.055}

\bibitem{PhysRevA.80.012102}
Cabello A, L\'opez-Tarrida A~J, Moreno P and Portillo J~R 2009 {\em Phys. Rev.
  A\/} {\bf 80}(1) 012102
  \urlprefix\url{https://link.aps.org/doi/10.1103/PhysRevA.80.012102}

\bibitem{Ji2010}
Ji Z, Chen J, Wei Z and Ying M 2010 {\em Quantum Inf. Comput.\/} {\bf 10}
  97--108 \urlprefix\url{https://doi.org/10.26421/QIC10.1-2-8}

\bibitem{Simon2025}
Claudet N and Perdrix S 2025 Local equivalence of stabilizer states: A
  graphical characterisation (Schloss Dagstuhl – Leibniz-Zentrum für
  Informatik)
  \urlprefix\url{https://drops.dagstuhl.de/entities/document/10.4230/LIPIcs.STACS.2025.27}

\bibitem{Burchardt2025}
Burchardt A, de~Jong J and Vandré L 2025 Algorithm to verify local equivalence
  of stabilizer states (\textit{Preprint} \eprint{2410.03961})
  \urlprefix\url{https://arxiv.org/abs/2410.03961}

\bibitem{claudet2025}
Claudet N and Perdrix S 2025 Deciding local unitary equivalence of graph states
  in quasi-polynomial time (\textit{Preprint} \eprint{2502.06566})
  \urlprefix\url{https://arxiv.org/abs/2502.06566}

\bibitem{Vandre2024Marginals}
Vandr\'e L, de~Jong J, Hahn F, Burchardt A, G\"uhne O and Pappa A 2024 {\em
  arXiv preprint quant-ph/2406.09956\/} Accepted for publication in
  \emph{Physical Review A} on 18~March~2025 (\textit{Preprint}
  \eprint{2406.09956})

\bibitem{Burchardt2025_2}
Burchardt A and Hahn F 2025 {\em Quantum\/} {\bf 9} 1720 ISSN 2521-327X
  \urlprefix\url{http://dx.doi.org/10.22331/q-2025-04-24-1720}

\bibitem{figure.lu}
\urlprefix\url{https://github.com/MarcinPlodzien/Graph-states}

\end{thebibliography}
\end{document}